\documentclass[aps,prb,10pt,twocolumn,superscriptaddress,longbibliography,floatfix]{revtex4-2}
\usepackage[T1]{fontenc}
\usepackage[utf8]{inputenc}
\usepackage{amsmath,amssymb,bm,graphicx,microtype}
\usepackage[colorlinks=true,linkcolor=blue,citecolor=blue,urlcolor=blue]{hyperref}

\usepackage{etoolbox}
\makeatletter
\patchcmd{\section}{0.8cm \@plus1ex \@minus .2ex}{9pt plus 1pt minus 1pt}{}{\PackageError{compact-layout}{Section spacing patch failed}{}}
\patchcmd{\section}{0.5cm}{4pt}{}{\PackageError{compact-layout}{Section spacing patch failed}{}}
\patchcmd{\subsection}{.8cm \@plus1ex \@minus .2ex}{8pt plus 1pt minus 1pt}{}{\PackageError{compact-layout}{Subsection spacing patch failed}{}}
\patchcmd{\subsection}{.5cm}{3pt}{}{\PackageError{compact-layout}{Subsection spacing patch failed}{}}
\patchcmd{\subsubsection}{.8cm \@plus1ex \@minus .2ex}{6pt plus 1pt minus 1pt}{}{\PackageError{compact-layout}{Subsubsection spacing patch failed}{}}
\patchcmd{\subsubsection}{.5cm}{4pt}{}{\PackageError{compact-layout}{Subsubsection spacing patch failed}{}}
\newcommand{\compactdisplays}{%
 \abovedisplayskip=4pt plus 1pt minus 1pt
 \belowdisplayskip=4pt plus 1pt minus 1pt
 \abovedisplayshortskip=1pt plus 1pt
 \belowdisplayshortskip=4pt plus 1pt minus 1pt}
\newcommand{\compactmathspacing}{%
 \thinmuskip=2.5mu
 \medmuskip=3mu plus 1mu minus 1mu
 \thickmuskip=4mu plus 1mu minus 1mu}
\AtBeginDocument{%
 \apptocmd{\normalsize}{\compactdisplays}{}{}
 \apptocmd{\small}{\compactdisplays}{}{}
 \compactdisplays
 \compactmathspacing
 \setlength{\parskip}{0pt}
 \raggedbottom
 \setlength{\textfloatsep}{8pt plus 1pt minus 1pt}
 \setlength{\floatsep}{8pt plus 2pt minus 2pt}
 \setlength{\intextsep}{8pt plus 2pt minus 2pt}
 \setlength{\abovecaptionskip}{4pt}
 \setlength{\belowcaptionskip}{0pt}
 \setlength{\jot}{1.5pt}}
\makeatother

\makeatletter
\DeclareRobustCommand{\bibtitlequote}[1]{``#1\@ifnextchar,{\bibtitlecomma}{''}}
\def\bibtitlecomma,{,''}
\makeatother
\newcommand{\tr}{\operatorname{Tr}}
\newcommand{\id}{\mathbb{I}}
\newcommand{\ket}[1]{\lvert#1\rangle}
\newcommand{\bra}[1]{\langle#1\rvert}
\newcommand{\norm}[1]{\left\lVert#1\right\rVert}
\newcommand{\cE}{\mathcal{E}}
\newcommand{\cM}{\mathcal{M}}
\newcommand{\cR}{\mathcal{R}}
\newcommand{\dd}{\mathrm{d}}
\newcommand{\ii}{\mathrm{i}}
\newcommand{\ee}{\mathrm{e}}
\newcommand{\Sswap}{\mathsf{S}}
\hypersetup{pdftitle={Partial projected ensembles reveal slow Stark-constrained information spreading},pdfauthor={Yi-Rui Zhang; Han-Ze Li; Jian-Xin Zhong},pdfsubject={PRB-style research manuscript; figure-based numerical analysis and exact analytical results}}
\begin{document}
\title{Partial projected ensembles reveal slow tilt-constrained information spreading}
\author{Yi-Rui Zhang}
\affiliation{Institute for Quantum Science and Technology, Shanghai University, Shanghai 200444, China}
\author{Yu-Jun Zhao}
\affiliation{School of Physics and Optoelectronics, Xiangtan University, Xiangtan 411105, China}
\affiliation{Institute for Quantum Science and Technology, Shanghai University, Shanghai 200444, China}

\author{Han-Ze Li}
\email{hanzeli@u.nus.edu}
\affiliation{Institute for Quantum Science and Technology, Shanghai University, Shanghai 200444, China}
\affiliation{Department of Physics, National University of Singapore, Singapore 117542, Singapore}
\author{Jian-Xin Zhong}
\email{jxzhong@shu.edu.cn}
\affiliation{Institute for Quantum Science and Technology, Shanghai University, Shanghai 200444, China}
\affiliation{School of Physics and Optoelectronics, Xiangtan University, Xiangtan 411105, China}
\begin{abstract}
Projected ensembles reveal information about quantum many-body dynamics beyond the reduced density matrix. Here we investigate how spatial constraints affect the emergence of this information in a kicked Ising chain with a spatially varying longitudinal field. We consider partial projected ensembles, obtained by measuring a remote region while leaving an intervening buffer unobserved. Our numerical results reveal a pronounced contrast between nearly ballistic correlation onset in the ergodic regime and strongly delayed onset in the tilt-constrained regime, corroborated by quantum mutual information. Despite these different onset scales, the conditional-state fluctuations decrease approximately exponentially with buffer length in both regimes. Full projected ensembles additionally exhibit slow relaxation and persistent measurement-basis dependence in their higher moments. We establish an information-theoretic characterization of the connected second moment and use a solvable dephasing model with effective-basis measurements to illustrate the separation of phase accumulation and coherence loss. Within this model, a factorially suppressed coupling envelope yields sublogarithmic spreading consistent with the finite-distance onset trends. Our results identify partial projected ensembles as probes of the distinction between the time required for correlations to develop and the information that remains accessible under partial observation.
\end{abstract}
\maketitle

\section{Introduction}
How quantum many-body dynamics redistributes information is central to understanding thermalization~\cite{Polkovnikov2011,Eisert2015,Reimann2008,Linden2009,Gogolin2016}. Canonical typicality~\cite{Popescu2006,Goldstein2006} and the eigenstate thermalization hypothesis~\cite{Deutsch1991,Srednicki1994,Rigol2008,DAlessio2016,Deutsch2018} explain how local equilibrium emerges at the level of reduced density matrices. Spatially resolved measurements in quantum simulators now probe a richer description by conditioning a subsystem on outcomes recorded in its environment~\cite{Choi2023Nature,Yan2026}. The resulting projected ensemble pairs conditional pure states with their Born probabilities: its first moment recovers the reduced density matrix, while higher moments retain fluctuations discarded by the partial trace~\cite{Cotler2023}. The emergence of universal conditional-state distributions in chaotic dynamics is known as deep thermalization~\cite{HoChoi2022,Cotler2023,Choi2023Nature,IppolitiHo2022,ClaeysLamacraft2022,IppolitiHo2023}. In the unconstrained infinite-temperature setting, the Haar ensemble and its finite-moment approximations, quantum state designs~\cite{Ambainis2007,Scott2006,Gross2007}, provide natural benchmarks.

\begin{figure}[!tbp]
\centering
\includegraphics[width=\columnwidth]{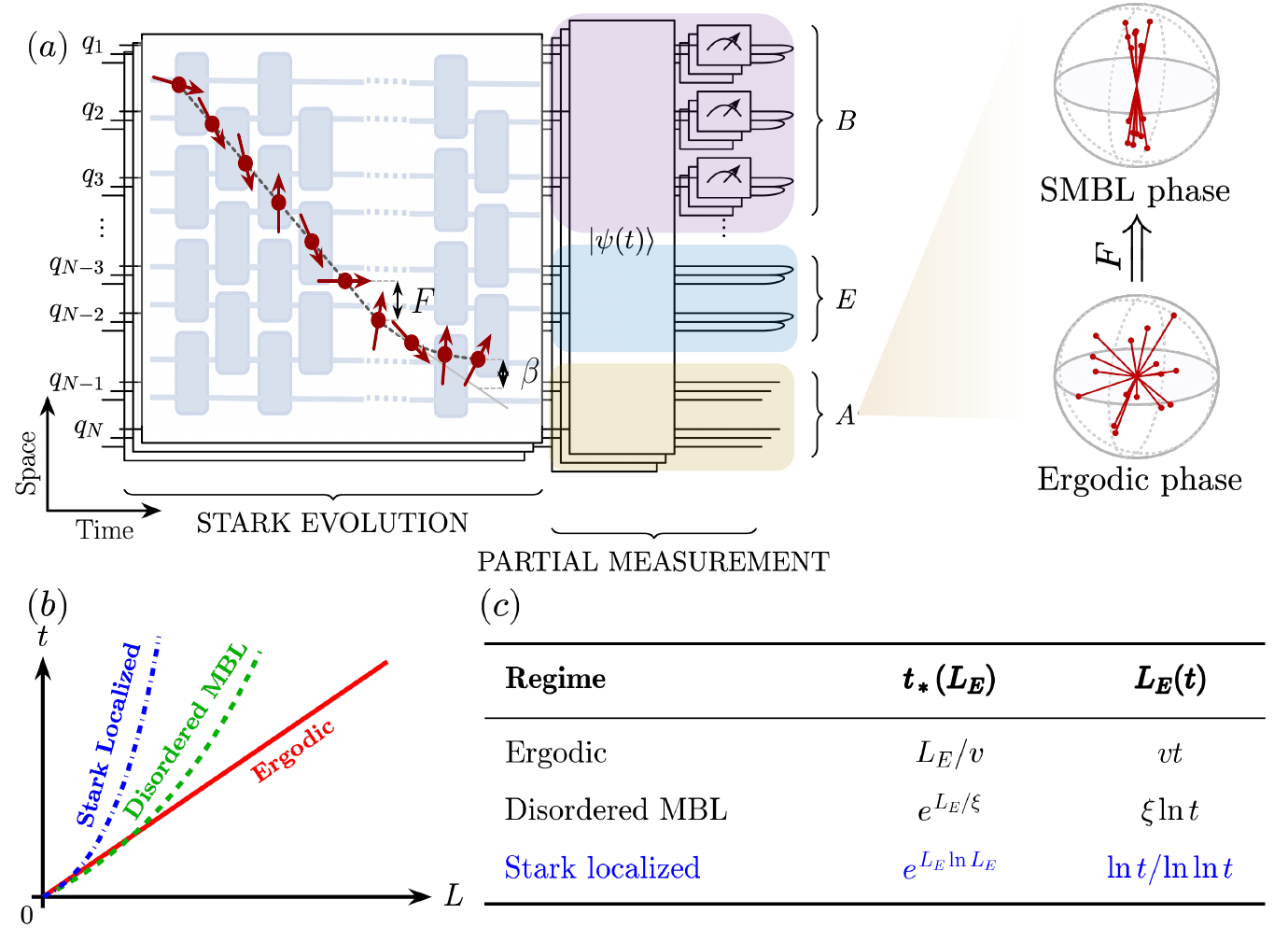}
\caption{Partial observation and proposed spreading laws. (a) Evolution in a Stark field with control parameter $F$ and curvature $\beta$, followed by measurement of $B$ and tracing out the buffer $E$, produces conditional states on $A$. Bloch spheres illustrate their distributions. (b),(c) Ballistic, logarithmic, and proposed sublogarithmic spreading correspond to linear, exponential, and factorial-type onset times versus buffer length $L_E$.}
\label{fig:geometry}
\end{figure}

These benchmarks change when physical constraints restrict the accessible states or the information revealed by measurements. Scrooge ensembles provide references with a prescribed, potentially nonmaximally mixed mean~\cite{Jozsa1994,Wootters2018,Mok2026}, supported by maximum-entropy principles for deep thermalization~\cite{Mark2024}. Symmetry and charge conservation make the limiting distribution sensitive to both initial-state fluctuations and the measurement basis~\cite{Varikuti2024,Chang2025}. Extensions to free-fermion and Gaussian systems~\cite{Lucas2023,Bejan2025,Liu2024Gaussian}, together with coherence-controlled ensemble transitions~\cite{Liu2026Coherence}, further establish that the reference ensemble and the observation protocol are essential to characterizing equilibration beyond its first moment.

The observation protocol becomes especially consequential when part of the environment remains unobserved, leaving conditional states mixed. Mixed-state deep thermalization connects this setting to universal random density-matrix ensembles~\cite{Yu2025,Zyczkowski2001,Sommers2004}. Partial projected ensembles (PPEs) give it a spatial interpretation: a region $B$ is measured, an intervening buffer $E$ is traced out, and the conditional states of a target $A$ are retained~\cite{Mandal2026} [Fig.~\ref{fig:geometry}(a)]. Varying the buffer probes the spatial development of dependence on distant outcomes and the information that remains accessible under partial observation~\cite{Mandal2026,Sherry2026}. Here measurements condition the final state; in measurement-induced entanglement transitions, they repeatedly modify the evolution~\cite{LiChenFisher2018,Skinner2019,Chan2019,Fisher2023,Li2025Measurement}.

This spatial perspective is particularly useful when information spreads despite suppressed transport. In disordered many-body localized systems~\cite{Basko2006,Oganesyan2007,Imbrie2016,Nandkishore2015,Abanin2019}, interactions between approximately local conserved degrees of freedom sustain logarithmic entanglement growth~\cite{Bardarson2012,Serbyn2013Growth,Serbyn2013,Huse2014}. A spatial tilt offers a setting for slow dynamics without quenched disorder: studies of Stark many-body localization~\cite{Schulz2019,Nieuwenburg2019,Doggen2021,Yao2021,Gunawardana2022} and experiments on tilted interacting systems~\cite{GuardadoSanchez2020,Scherg2021,Morong2021,Guo2021} reveal suppressed transport and persistent local memory. Slow growth of stabilizer R\'enyi entropy provides a complementary probe beyond conventional entanglement~\cite{Leone2022,Li2025}. In periodically driven systems, however, long prethermal regimes and the stability of localization require distinguishing finite-time slowing from asymptotic behavior~\cite{DAbanin2015Heating,Kuwahara2016,DAbanin2017Rigorous,Ho2023Review,Sierant2023,Duffin2024,Zisling2022}. Local memory therefore leaves two questions open: how quickly do distant measurement outcomes become correlated with a subsystem, and how much of that correlation survives when the intervening region is unobserved? Conditional phase accumulation and the loss of coherence upon tracing out the buffer suggest that these questions probe distinct aspects of the dynamics.

In this work, we address these questions using projected ensembles in a kicked Ising chain~\cite{Bertini2018} with a spatially varying longitudinal field. We find strongly delayed correlation growth in the Stark-constrained regime, together with an approximately exponential reduction of conditional-state fluctuations with buffer length in both constrained and ergodic regimes. Numerical results and a solvable dephasing picture connect these observations to the distinct roles of phase accumulation and coherence loss. PPEs thus resolve both the time required for correlations to develop and the information retained across an unobserved region. Sections~\ref{sec:framework}--\ref{sec:mechanism} present the framework, results, and mechanism; Sec.~\ref{sec:conclusion} concludes, with technical details in the appendices.


\section{Settings}
\label{sec:framework}
\subsection{Geometry and dynamical setting}
We consider an open spin-$1/2$ chain partitioned into contiguous regions $A$, $E$, and $B$, with the target $A$ at one boundary and the unobserved buffer $E$ separating it from the measured region $B$ [Fig.~\ref{fig:geometry}(a)]. Their lengths are $L_A,L_E,L_B$, and their Hilbert-space dimensions are $d=2^{L_A}$, $e=2^{L_E}$, and $b=2^{L_B}$.

The dynamics is generated by a kicked Ising chain with a
weakly curved Stark field. One Floquet period is
\begin{equation}
 U_F=\ee^{-\ii K}\ee^{-\ii H_z}.
 \label{eq:floquet}
\end{equation}
The diagonal generator and transverse kick are
\begin{align}
 H_z&=J\sum_{j=0}^{L-2}Z_jZ_{j+1}
      +\sum_{j=0}^{L-1}h_jZ_j, \label{eq:hz}\\
 K&=g\sum_{j=0}^{L-1}X_j. \label{eq:kick}
\end{align}
Here $X_j,Z_j$ are Pauli operators, $\hbar=1$, and time
is measured in Floquet periods, with
$\ket{\Psi(t)}=U_F^t\ket{\Psi_0}$ at integer $t$.
The kick amplitude and field profile are parametrized by
$\Gamma\in[0,1]$ as
$g=g_0\Gamma$ and
$h_j=h_0+A_0\sqrt{1-\Gamma^2}[j+\beta(j/L)^2]$.
We use $J=0.85$, $h_0=0.63$, $g_0=0.70$, $A_0=0.55$,
and $\beta=0.1$, where $\beta$ specifies the quadratic
curvature of the field profile.
In the following, we use the reparametrized control parameter
$F=1-\Gamma$, with larger $F$ corresponding to a weaker
transverse kick and a stronger Stark tilt.
We compare $F=0.2$ and $F=0.8$ as representative ergodic
and Stark-constrained dynamical regimes, respectively.
The broader kicked Ising circuit family admits a
dual-unitary limit, although the parameters used here
are away from that solvable limit.
The transverse kick does not conserve total $Z$
magnetization.

\begin{figure}[!tbp]
\centering
\includegraphics[width=\columnwidth]{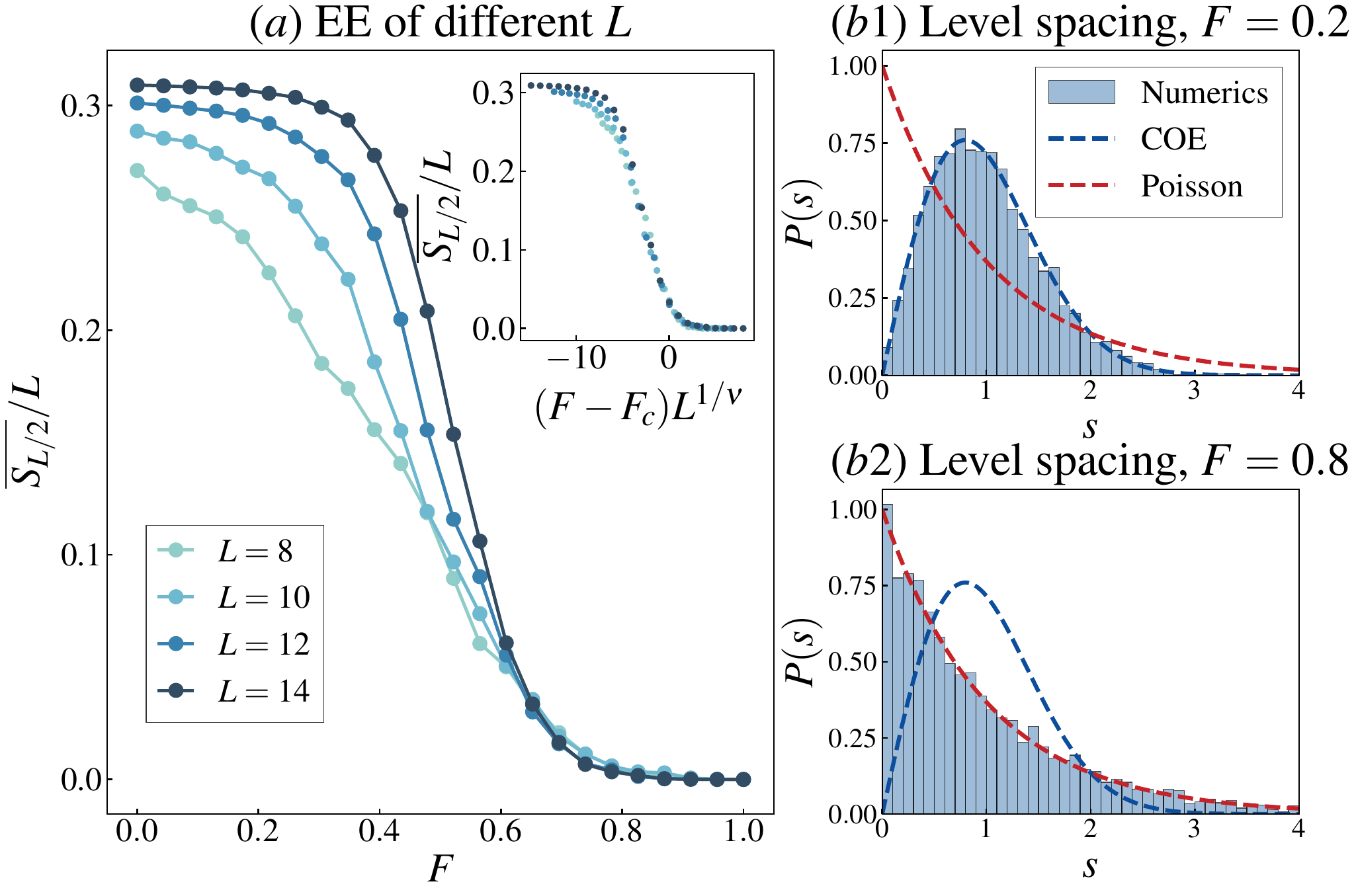}
\caption{Regime diagnostics. (a) Half-chain von Neumann entropy, averaged over all Floquet eigenstates and divided by $L$, versus control parameter $F$. Inset: rescaling with $F_c=0.651$ and $\nu=0.84$. (b1),(b2) Normalized circular quasienergy-spacing distributions for $L=12$ at $F=0.2$ and $0.8$; dashed curves show circular-orthogonal-ensemble (COE) and Poisson benchmarks. Entropies use natural logarithms.}
\label{fig:regimes}
\end{figure}

The ensemble dynamics is averaged over $N_{\mathrm{init}}=1000$ independent random product states $\ket{\Psi_0}=\bigotimes_{j=0}^{L-1}\ket{\psi_j}$. Each local state is $\ket{\psi_j}=\cos(\theta_j/2)\ket0_j+\ee^{\ii\phi_j}\sin(\theta_j/2)\ket1_j$, with $\cos\theta_j$ uniform on $[-1,1]$ and $\phi_j$ uniform on $[0,2\pi)$, independently at each site. Each diagnostic is evaluated for an individual initial state before averaging. Full-ensemble trajectories are evaluated from the Floquet eigendecomposition at logarithmically sampled integer times, whereas the PPE and mutual-information trajectories use successive applications of $U_F$. Numerical evaluation is described in Appx.~\ref{app:numerics}; averaging windows and conventions are specified in Appx.~\ref{app:provenance}.

We denote the measurement angle by $\alpha$. The local basis vectors are $\ket{0;\alpha}=\cos(\alpha/2)\ket0+\sin(\alpha/2)\ket1$ and $\ket{1;\alpha}=\sin(\alpha/2)\ket0-\cos(\alpha/2)\ket1$. Their tensor products define $\{\ket{z;\alpha}\}$ on the measured region. The choices $\alpha=0$ and $\pi/2$ give the $Z$ and $X$ bases. The full-ensemble basis comparison uses $\alpha/\pi=0.01,0.05,0.20,0.50$; all PPE results use the $X$ basis. The full ensembles have $L_A=1,2$ and $L=6$--$12$ at $F=0.8$. For spatially resolved PPE dynamics, we fix $L_A=1$ and $L_B=6$ while varying $L_E=1$--$9$, so that $L=L_E+7$. The mutual-information comparison uses the same geometry with $L_E=3$--$9$.

\subsection{Full projected ensembles}
For the full ensemble, $E$ is absent and $B$ is the entire complement of $A$. The unnormalized conditional vector after measuring $B$ is $\ket{\widetilde\psi_z}=\bra{z;\alpha}\Psi(t)\rangle$. Its squared norm gives the Born probability $p_z$, and the normalized state is $\ket{\psi_z}=\ket{\widetilde\psi_z}/\sqrt{p_z}$ for $p_z>0$. The full projected ensemble $\cE_A=\{p_z,\ket{\psi_z}\}$ has moments
\begin{equation}
 \cM_A^{(k)}=\sum_z p_z
       \big(\ket{\psi_z}\bra{\psi_z}\big)^{\otimes k}.
 \label{eq:fullmoment}
\end{equation}
We compare a moment with a specified reference $\cR_A^{(k)}$ through
\begin{equation}
 D_k=\frac12\norm{\cM_A^{(k)}-\cR_A^{(k)}}_1.
 \label{eq:Dk}
\end{equation}
For the full-ensemble data, $\cR_A^{(k)}$ is the Scrooge-ensemble moment constructed from the reduced Floquet diagonal ensemble. Writing $U_F\ket{\varphi_\mu}=\ee^{\ii\omega_\mu}\ket{\varphi_\mu}$ and $c_\mu=\langle\varphi_\mu|\Psi_0\rangle$, its mean is $\rho_{A,\infty}=\sum_\mu |c_\mu|^2 \tr_B\!\left(\ket{\varphi_\mu}\bra{\varphi_\mu}\right)$. The reference is constructed separately for each initial state and held fixed during its evolution. Appx.~\ref{app:full} gives the Scrooge construction. The plotted $\Delta^{(k)}$ is $D_k$ with this reference, averaged over the random product states.

For comparison, the Haar reference underlying quantum designs is $\cR_{\mathrm H}^{(k)}=\Pi_{\mathrm{sym}}^{(k)}/\binom{d+k-1}{k}$, where $\Pi_{\mathrm{sym}}^{(k)}$ projects onto the symmetric subspace. The Scrooge reference reduces to this expression when $\rho_{A,\infty}=\id/d$. In general, relaxation toward the specified Scrooge reference and formation of a Haar design are distinct questions.

Irrespective of measurement basis, $\cM_A^{(1)}=\rho_A$. Thus basis dependence of higher moments reveals ensemble information beyond the mean~\cite{BasisConsistencyNote}. For any consistent hierarchy of reference moments, partial-trace contractivity of trace distance~\cite{Fuchs1999} further gives $D_k\geq D_1$; a nonzero first-moment floor cannot be removed by increasing the moment order. Appx.~\ref{app:full} develops these constraints.

\subsection{Partial ensembles as conditional-state fluctuations}
With $E$ unobserved, measuring $B$ leaves the unnormalized vector $\ket{\widetilde\psi_z}_{AE}=\bra{z;\alpha}\Psi\rangle$. Tracing out the buffer gives $\rho_{A|z}=\tr_E(\ket{\widetilde\psi_z}\bra{\widetilde\psi_z})/p_z$, with $p_z=\langle\widetilde\psi_z|\widetilde\psi_z\rangle$ and $\rho_A=\sum_zp_z\rho_{A|z}$. The second PPE moment is $\cM_{A,\mathrm P}^{(2)}=\sum_zp_z\rho_{A|z}^{\otimes2}$. Its connected part is $C_A^{(2)}=\cM_{A,\mathrm P}^{(2)}-\rho_A^{\otimes2}$, and the associated diagnostic is
\begin{equation}
 \Delta=\frac12\norm{C_A^{(2)}}_1.
 \label{eq:Delta}
\end{equation}
This is not the full-ensemble distance in Eq.~\eqref{eq:Dk}. It tests how much the conditional states fluctuate around their own mean. With $\delta\rho_z=\rho_{A|z}-\rho_A$, one obtains the exact identity $C_A^{(2)}=\sum_zp_z\,\delta\rho_z\otimes\delta\rho_z$. Although $C_A^{(2)}$ need not be positive as an operator on two copies, it is a covariance tensor. For observables $O,O'$ on $A$, contracting it with $O\otimes O'$ yields the classical covariance of their conditional expectation values.

A particularly useful scalar is the conditional-state variance $V=\sum_zp_z\tr(\delta\rho_z^2) =\tr\big[\Sswap_A C_A^{(2)}\big]$, where $\Sswap_A$ swaps the two copies. It obeys $\frac{V}{2}\leq\Delta\leq\frac{dV}{2}$. Consequently, $\Delta=0$ if and only if every conditional state of nonzero probability equals $\rho_A$. An increasing $\Delta(t,L_E)$ signals the appearance of dependence on the distant measurement record, not a change in the unconditioned state caused by that measurement.

This interpretation can be made information theoretic. Let $Z$ denote the classical outcome register and define, with natural logarithms, $\chi(A{:}Z)=S(\rho_A)-\sum_zp_zS(\rho_{A|z})$. The quantity $\chi$ is the mutual information of the classical--quantum state $\sum_zp_z\rho_{A|z}\otimes\ket{z}\bra{z}$, often called its Holevo information~\cite{Holevo1973}. The second spatial diagnostic is the premeasurement quantum mutual information $I_q(A{:}B)=S(\rho_A)+S(\rho_B)-S(\rho_{AB})$. Since the global state is pure, $S(\rho_{AB})=S(\rho_E)$, which is the form used numerically. These quantities satisfy
\begin{equation}
 \Delta\leq\chi(A{:}Z)\leq I_q(A{:}B).
 \label{eq:infobounds}
\end{equation}
The detailed proof is given in Appx.~\ref{app:cov}. A nonzero premeasurement mutual information need not be visible in a particular measurement basis. Thus the two diagnostics probe related but distinct correlations. Their onset curves use the same absolute threshold, $10^{-5}$, applied separately to the initial-state-averaged $\Delta$ and $I_q$.


\section{Phenomenology}
\label{sec:results}
\subsection{Regime identification and full projected ensembles}
Fig.~\ref{fig:regimes} identifies the parameter regimes used in the following dynamical comparisons. The normalized half-chain entropy decreases strongly with the control parameter $F$ for $L=8,10,12,14$. At $F=0.2$, the relatively large eigenstate entanglement and level repulsion in the quasienergy-spacing distribution~\cite{Oganesyan2007,Atas2013} are characteristic of ergodic dynamics. At $F=0.8$, the reduced entanglement and approximately Poisson spacing statistics indicate the Stark many-body-localized regime. We therefore choose $F=0.2$ and $F=0.8$ as representative ergodic and Stark-constrained parameters, respectively, for comparing projected-ensemble relaxation and spatial information spreading. Additional results are provided in Appx.~\ref{app:supporting}.

\begin{figure}[!tbp]
\centering
\includegraphics[width=\columnwidth]{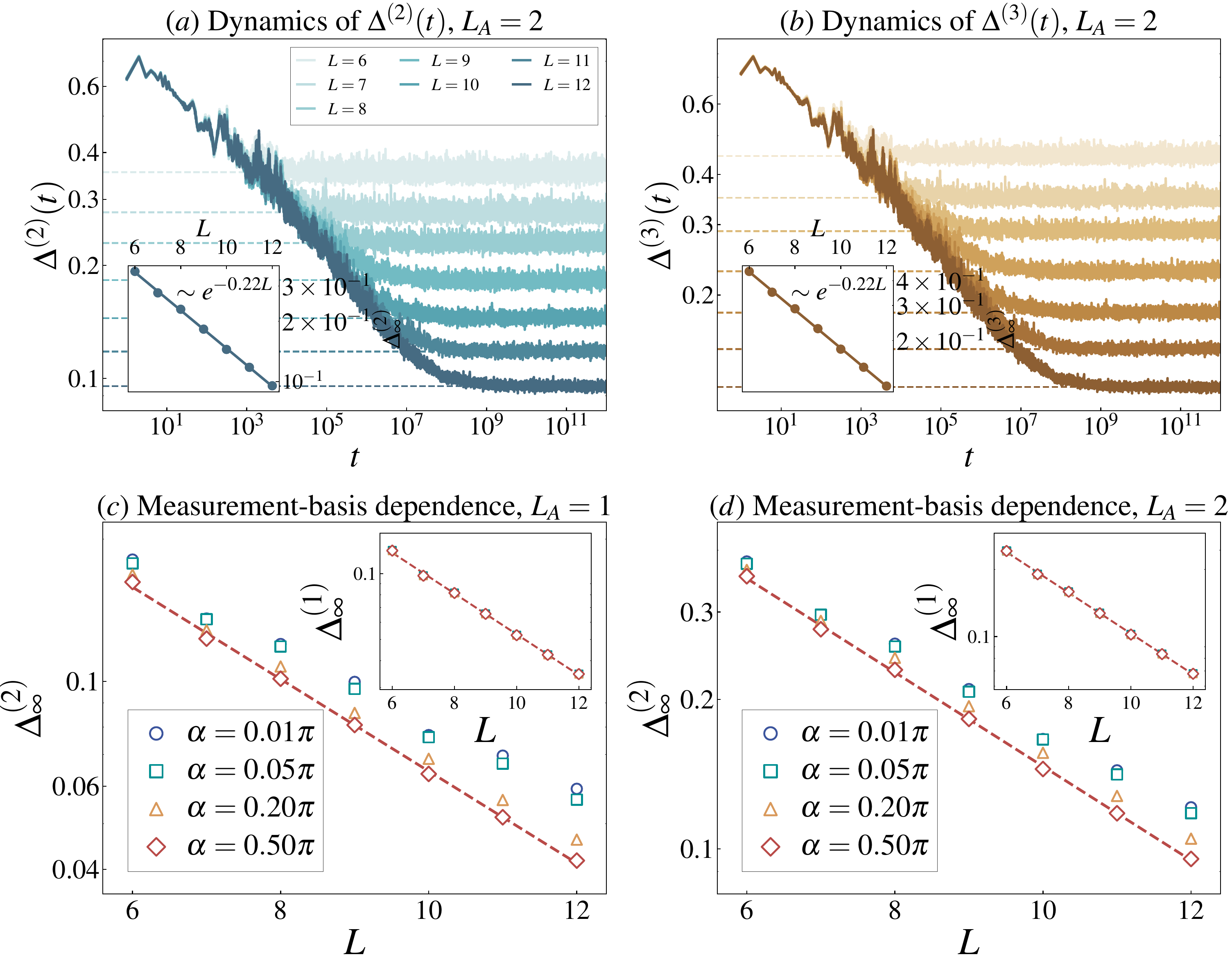}
\caption{Full projected ensembles at $F=0.8$, averaged over 1000 random product states; $t$ counts Floquet periods. (a),(b) Second- and third-moment trace distances $\Delta^{(k)}$ to the initial-state-specific Scrooge reference, with $L_A=2$ and $X$-basis measurements. (c),(d) Residual second-moment distances versus $L$ and measurement angle $\alpha$ for $L_A=1,2$; insets show first moments. Here $\alpha=\pi/2$ is the $X$ basis. Residuals and horizontal dashed lines use the last 200 saved times; inset lines in (a),(b) and dashed curves in (c),(d) are exponential fits.}
\label{fig:full}
\end{figure}

Fig.~\ref{fig:full}(a),(b) shows slow relaxation of the second- and third-moment distances to the Scrooge reference for $L_A=2$, extending over many decades in time. Increasing the total system size reduces the residual distances, whose averages over the last 200 saved times scale approximately as $e^{-0.22L}$ for both orders over $L=6$--$12$. The second- and third-moment results for $L_A=1$ and the
first-moment results for $L_A=1,2$ in
Appx.~\ref{app:supporting} exhibit comparable size dependence. Thus, slow temporal relaxation coexists with a systematic reduction of the residual moment distances as the measured environment grows.

The measurement-angle dependence in Fig.~\ref{fig:full}(c),(d) reveals structure beyond the reduced density matrix. The residual second-moment distance varies substantially with $\alpha$, while the first-moment curves in the insets are independent of the measurement basis. This follows from the ensemble construction: changing which environmental observable is measured changes the conditional states and their probabilities, while their average remains $\rho_A$. The basis dependence therefore resolves anisotropy in the conditional-state distribution that is absent from its first moment.

The moment hierarchy helps interpret these trends. Since $D_k\geq D_1$ for the consistent reference moments used here, relaxation of the reduced state also contributes to the higher-moment distances. Dephasing can nevertheless leave nontrivial higher-moment structure: the equatorial qubit ensemble in Appx.~\ref{app:full} has mean $\id/2$ but a second-moment Haar distance of $1/6$. These features connect the observed basis dependence to the distribution of conditional states. Accordingly, we describe Fig.~\ref{fig:full} as slow projected-ensemble relaxation.

\subsection{Partial ensembles and spatial information spreading}
\begin{figure}[!tbp]
\centering
\includegraphics[width=\columnwidth]{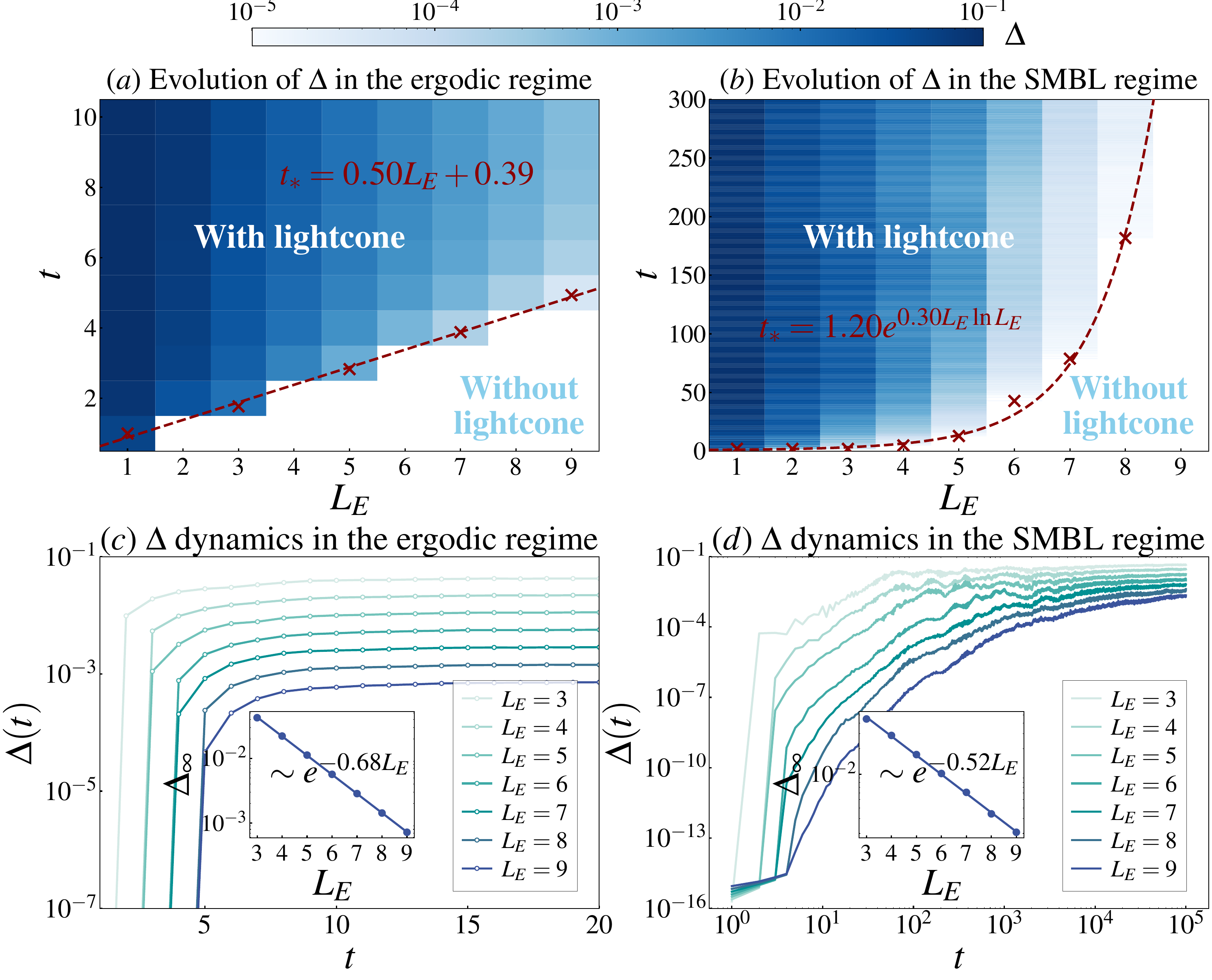}
\caption{PPE connected second-moment trace distance $\Delta$ for $L_A=1$, $L_B=6$, $L=L_E+7$, and $X$-basis measurement of $B$, averaged over 1000 random product states. (a),(c) $F=0.2$; (b),(d) $F=0.8$. Heatmaps show $\Delta(t,L_E)$; crosses mark the $10^{-5}$ onset threshold and red dashed curves are the displayed fits. The light-cone labels distinguish the two sides of this threshold. Insets in (c),(d) show residuals at $t=20$ and averaged over the last 200 saved times, respectively, with exponential fits. Time is in Floquet periods.}
\label{fig:ppe}
\end{figure}

The PPE resolves the spatial component of this dynamics by separating $A$ from the measured region $B$ with an unobserved buffer of length $L_E$. In Fig.~\ref{fig:ppe}, the connected second-moment distance $\Delta$ grows after a delay that increases with the buffer length. The contour at $\Delta=10^{-5}$ defines the onset time $t_*$ and separates the low-signal region from the region in which conditional-state fluctuations become resolved. This common threshold makes the contrast between ergodic and Stark-constrained dynamics directly visible.

In the ergodic regime, the onset follows
\begin{equation}
 t_*^{\mathrm{erg}}(L_E)\simeq0.50L_E+0.39.
 \label{eq:ergfit}
\end{equation}
The approximately linear growth is consistent with ballistic correlation spreading: each additional buffer site adds a nearly constant delay. The slope measures the onset of equal-time correlations under the specified measurement and threshold protocol, while the intercept accounts for the short-distance delay.

The Stark-constrained regime instead exhibits a strongly nonlinear increase in onset time, described over the sampled separations by
\begin{equation}
 t_*^{\mathrm{St}}(L_E)\simeq
       1.20e^{0.30L_E\ln L_E}.
 \label{eq:stfit}
\end{equation}
The contrast with Eq.~\eqref{eq:ergfit} demonstrates a pronounced delay in correlation buildup in the Stark-constrained regime relative to the ergodic regime. The $L_E\ln L_E$ dependence is consistent with the factorial-type suppression of effective couplings developed in Sec.~\ref{sec:mechanism}. It provides a description of the slow onset and motivates the corresponding sublogarithmic spreading picture.

The signal amplitude and onset time probe different aspects of the dynamics. The ergodic values at $t=20$ decrease approximately as $\Delta(20,L_E)\propto e^{-0.68L_E}$, while the constrained values averaged over the last 200 saved times follow $\overline{\Delta}(L_E)\propto e^{-0.52L_E}$. Both show exponential attenuation as more spins are left unobserved, although their onset times have very different distance dependences. Tracing out the buffer reduces the conditional information accessible on $A$; the onset contour separately records the time required for that information to develop.

An induced-state ensemble gives a useful reference for this attenuation. Tracing an $e$-dimensional environment from an isotropic pure-state ensemble yields
\begin{equation}
 \Delta_{\mathrm{ind}}=
       \frac{d^2-1}{2d(de+1)}.
 \label{eq:inducedmain}
\end{equation}
For fixed $d$ and a qubit buffer, the resulting $2^{-L_E}$ scaling has an exponent $\ln2\simeq0.693$, close to the ergodic value $0.68$. This benchmark illustrates the suppression of conditional-state fluctuations by averaging over unobserved degrees of freedom. Appx.~\ref{app:random} derives the finite-record correction and distinguishes the norm of the averaged covariance from the average norm used in the numerical diagnostic.

\begin{figure}[!tbp]
\centering
\includegraphics[width=\columnwidth]{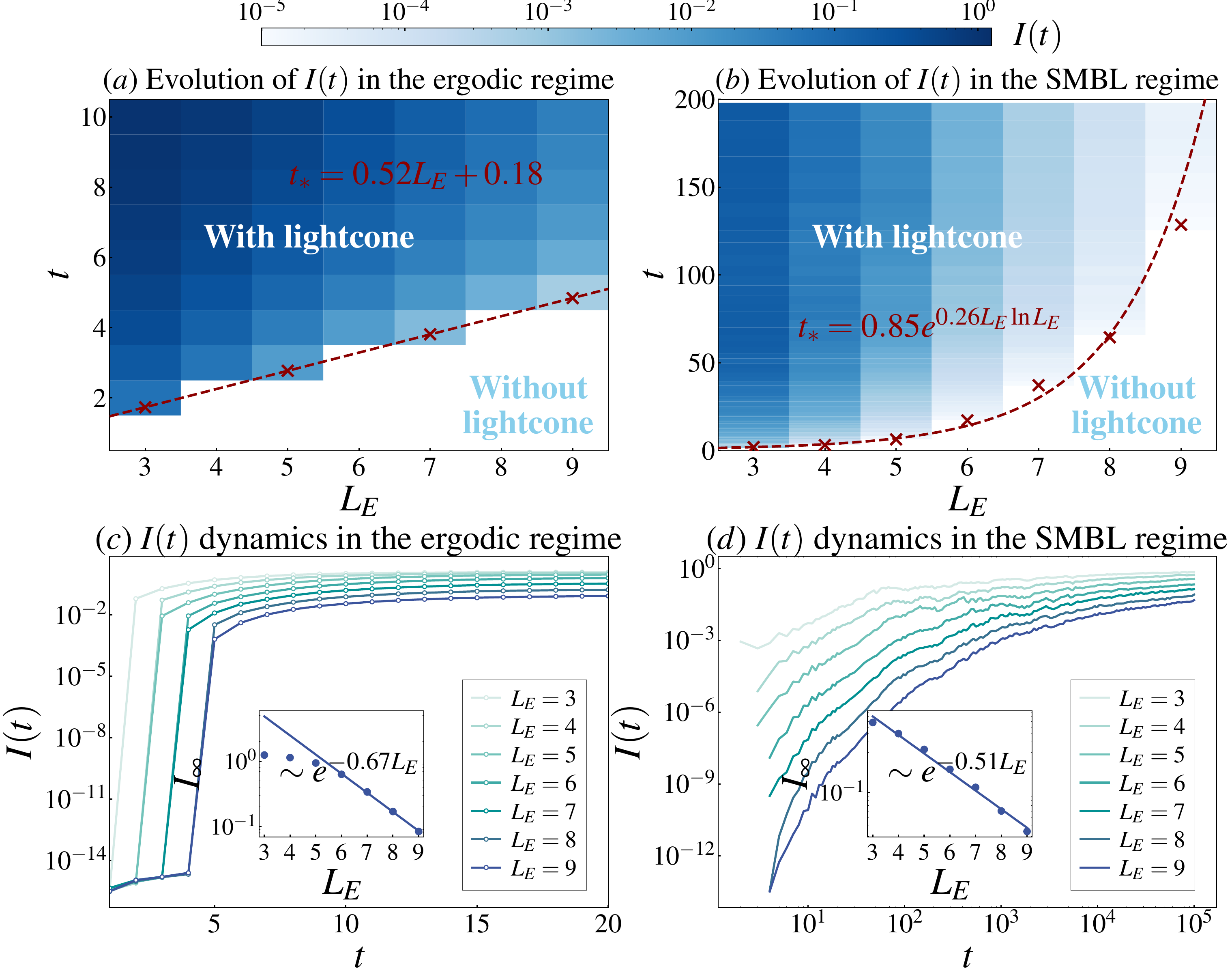}
\caption{Premeasurement quantum mutual information $I=S_A+S_B-S_{AB}$, using natural logarithms, with $L_A=1$, $L_B=6$, and $L=L_E+7$. Results average 1000 random product states; $t$ counts Floquet periods. (a),(c) $F=0.2$; (b),(d) $F=0.8$. Crosses mark $I=10^{-5}$ and red dashed curves are onset fits. Insets show $I(t=20)$ in (c) and the mean over  the last 20 saved times in (d), with exponential fits. The $0.67$ fit uses $L_E=6$--$9$; the $0.51$ fit uses $L_E=3$--$9$ and describes a finite-time average.}
\label{fig:mi}
\end{figure}

The premeasurement quantum mutual information in Fig.~\ref{fig:mi} independently resolves the same spatial contrast. Its onset fits are
\begin{align}
 t_{*,I}^{\mathrm{erg}}&\simeq0.52L_E+0.18,\\
 t_{*,I}^{\mathrm{St}}&\simeq0.85e^{0.26L_E\ln L_E}.
 \label{eq:mifits}
\end{align}
The nearly linear ergodic onset and strongly delayed Stark-constrained onset thus appear both in the correlations of the premeasurement state and in the conditional-state fluctuations of the PPE. The mutual-information amplitudes also decrease exponentially with buffer length: the values at $t=20$ give $e^{-0.67L_E}$ in the ergodic regime, while the average over the last 20 saved times gives $e^{-0.51L_E}$ in the Stark-constrained regime.

These diagnostics access complementary information. The quantum mutual information $I_q$ includes correlations present before measurement, whereas the PPE probes the conditional information retained by the chosen $X$-basis measurement. Their magnitudes obey Eq.~\eqref{eq:infobounds}, and their different sensitivities lead to different threshold-crossing times and fitted coefficients. Their common distance dependence identifies slow correlation buildup as the physical origin of the delayed PPE response and motivates the effective dephasing picture discussed next.

\section{Discussion}
\label{sec:mechanism}
The numerical results distinguish the time required for information to develop across an unobserved region from the amount of information that survives partial observation. At $F=0.8$, the full projected ensemble relaxes slowly and retains measurement-basis dependence in its higher moments. The PPE adds a spatial perspective: increasing the buffer length strongly delays its response while reducing its amplitude. The mutual-information results show a similar delay before measurement. Together, these observations motivate a description in terms of weak interactions that generate correlations through conditional phase accumulation.

A useful effective model is the diagonal Hamiltonian $H_{\mathrm d}=\sum_i h_i\tau_i^z+\sum_{i<j}J_{ij}\tau_i^z\tau_j^z$. Here $\tau_i^z=\mathcal U Z_i\mathcal U^\dagger$ is an effective Pauli operator with eigenvalues $\pm1$, obtained from the physical spin by a quasi-local unitary dressing transformation $\mathcal U$. It is centered near site $i$ but generally contains multispin contributions extending to neighboring sites, so it differs from the bare operator $Z_i$. These dressed spins, often called l-bits, satisfy $[H_{\mathrm d},\tau_i^z]=0$: their longitudinal populations are conserved in the effective model, while their coherences can evolve. For the original Floquet chain, we use this description only on finite-time scales where nonresonant dressing and the pairwise truncation remain valid. The construction and its assumptions are explained in Appx.~\ref{app:dephasing}. In this description, weak couplings let distant spins change the phase of a local superposition while local populations retain memory. Measurement in the effective longitudinal basis resolves the configurations generating these phases, whereas tracing out spins averages over them. This solvable setting illustrates how correlation buildup and loss of accessible coherence can occur on different scales.

This separation is explicit for one target effective spin, an initial product state, and measurement of $B$ in the effective-spin basis. Let $q_j$ be the initial probability of $\tau_j^z=+1$ and $c_0$ the initial target coherence. The conditional coherence is
\begin{equation}
 c_z(t)=c_0\ee^{-2\ii h_At}G_E(t)
 \ee^{-2\ii t\sum_{j\in B}J_{Aj}z_j},
 \label{eq:discussioncoherence}
\end{equation}
where $G_X(t)=\prod_{j\in X}[q_j\ee^{-2\ii J_{Aj}t}+(1-q_j)\ee^{2\ii J_{Aj}t}]$ for $X=E,B$. All outcomes share the magnitude $|c_0G_E|$, while the recorded spins determine their relative phases. Interactions entirely within the environment cancel from the conditional reduced state. The numerical PPE instead uses physical $X$-basis measurements. Those projectors and the initial state are transformed by dressing, so the factorization is an illustrative effective-basis result, not an identity for the numerical PPE. Appx.~\ref{app:generalmeasurement} gives the general measurement expression and its relation to physical projectors; the covariance identities and information bounds in Appx.~\ref{app:cov} hold for either measurement choice.

The conditional-state variance is
\begin{equation}
 V(t)=2|c_0|^2|G_E(t)|^2[1-|G_B(t)|^2].
 \label{eq:discussionvariance}
\end{equation} Here $1-|G_B|^2$ measures the spread of record-dependent phases, and $|G_E|^2$ measures the remaining coherence after the buffer is discarded. The trace-distance diagnostic takes the analogous form
\begin{align}
 \Delta(t)&=|c_0|^2|G_E(t)|^2\mathcal F_B(t),
 \label{eq:discussiondistance}\\
 \mathcal F_B(t)&=1-|G_B(t)|^2
 +|G_B^{(2)}(t)-G_B(t)^2|,
 \label{eq:discussionphasefactor}
\end{align}
where $G_B^{(2)}=\prod_{j\in B}[q_j\ee^{-4\ii J_{Aj}t}+(1-q_j)\ee^{4\ii J_{Aj}t}]$. The second harmonic retains anisotropy of the phase distribution. The exact relations and their derivations are collected in Appx.~\ref{app:dephasing}.

The buffer factor provides a simple interpretation of exponential signal attenuation. For nonresonant buffer frequencies, its long-time average is $\overline{|G_E|^2}=\prod_{j\in E}[q_j^2+(1-q_j)^2]$. Each ignored spin contributes a coherence-reduction factor: balanced populations give $2^{-L_E}$, and biased populations give weaker attenuation. When the measured spins generate a broad phase distribution, $\mathcal F_B$ is typically of order unity, so the buffer product sets the remaining signal scale. Within this model, approximately exponential amplitudes can therefore coexist with different propagation laws. The buffer product does not determine the decay slopes measured in the physical $X$ basis.

In the solvable effective-basis setting, the phase-spreading time is controlled by the width of the remote conditional field $H_B=\sum_{j\in B}J_{Aj}z_j$. For independent initial spins, its variance is $\sigma_B^2=4\sum_{j\in B}q_j(1-q_j)J_{Aj}^2$. The expansion $1-|G_B|^2=4t^2\sigma_B^2+O(t^4)$ shows that narrower field distributions require longer times to produce distinguishable phases. A fixed normalized response therefore gives $t_*\sigma_B\sim1$. If the nearest active couplings dominate, this becomes $t_*(r)J_{\mathrm{eff}}(r)\sim1$, with $r=L_E+O(1)$. Absolute-threshold onsets also depend on signal attenuation~\cite{ThresholdOnsetNote}. Appx.~\ref{app:fronts} details the relation between intrinsic phase-spreading scales and threshold-crossing times.

A static tilt suggests how remote couplings can decrease faster than exponentially. If successive virtual steps have detunings $\mathcal E,2\mathcal E,\ldots,(r-1)\mathcal E$ and elementary matrix elements of order $g$, the amplitude scales as $A_r\sim g(g/\mathcal E)^{r-1}/(r-1)!$. Here $\mathcal E$ is an energy increment, distinct from the control parameter $F$ of the kicked chain. Each extra step increases both the process order and its detuning. The resulting factorial suppression motivates the effective envelope $J_{\mathrm{eff}}(r)=J_0\exp[-ar\ln r-b_1r+o(r)]$, with $a>0$. Its coefficients encode the dominant interacting paths and their interference.

In a Floquet system, the relevant denominators are quasienergy differences of magnitude $2|\sin[T(\varepsilon_m-\varepsilon_n)/2]|$. They reproduce unfolded energy differences times $T$ within a nonresonant small-phase window, but become periodic at larger phase separations. Accordingly, the factorial envelope is an effective hypothesis for distances and times dominated by nonresonant processes. The dephasing solution specifies the response to such an envelope, while its microscopic realization depends on the dressed couplings of the kicked chain. The construction and denominator structure are given in Appx.~\ref{app:dephasing} and Appx.~\ref{app:fronts}.

Under this hypothesis, the onset obeys $\ln[t_*(r)/t_0]=ar\ln r+b_1r+o(r)$. The fitted constrained onsets have this form with the linear correction omitted and are consistent with it over the sampled separations. Retaining the linear term and setting $y=\ln(t/t_0)$ gives $r(t)=y/\{aW_0[(y/a)\ee^{b_1/a}]\}$, where $W_0$ is the principal Lambert function~\cite{Corless1996}. The large-time prediction is $r(t)\sim\ln(t/t_0)/[a\ln\ln(t/t_0)]$, a sublogarithmic front. The inversion and finite-distance corrections are derived in Appx.~\ref{app:fronts}. This connects the proposed coupling envelope to the observed delayed onset while distinguishing the asymptotic model prediction from the finite-distance numerical fits.

The resulting physical picture separates two effects of partial observation. Weak remote couplings set the time required for different measurement records to generate distinguishable phases; unobserved spins set how much coherence survives to reveal those differences. A rapidly increasing onset time can therefore accompany an exponentially decreasing response. This separation provides a qualitative interpretation of the numerical trends. The shared delay of the physical $X$-basis PPE and premeasurement mutual information supports slow correlation buildup; attributing its detailed measurement response to the factorized dephasing mechanism additionally requires control of the dressed state and projectors.


\section{Conclusion}
\label{sec:conclusion}
In this work, we used projected ensembles to investigate how information becomes accessible under partial observation in a kicked Ising chain with a spatially varying longitudinal field. In the Stark-constrained regime, full projected ensembles relax slowly toward the Scrooge reference, with residual moment distances decreasing as the system grows. Their higher moments retain measurement-basis dependence even though the average state is basis independent, revealing conditional-state structure beyond the reduced density matrix. Partial projected ensembles resolve the spatial buildup of this information: correlation onset grows approximately linearly with buffer length in the ergodic regime, but is strongly delayed in the Stark-constrained regime, where the finite-distance fits are consistent with $\ln t_*\propto L_E\ln L_E$. Quantum mutual information corroborates this contrast, while the PPE signal decreases approximately exponentially with buffer length in both regimes. We identify the connected second moment as a covariance of conditional states and bound its trace-distance diagnostic by the record's Holevo information and the premeasurement mutual information. A solvable dephasing model with effective-basis measurements illustrates how weak remote couplings delay conditional phase spreading while unobserved spins attenuate coherence. Within this model, a factorial-type coupling envelope yields sublogarithmic spreading, suggesting a physical interpretation of the observed onset trends. Together, these results establish PPEs as probes that distinguish the time required for correlations to develop from the information retained after partial observation.

Larger systems and microscopic estimates of effective couplings, together with the dressed physical measurement operators, could clarify the range and mechanism of the observed spreading law. On quantum hardware, conditional measurements could test the predicted separation of onset time and surviving signal by varying the buffer and measurement basis. Appx.~\ref{app:implementation} outlines the implementation and estimation requirements. These directions extend PPEs as probes of how spatial constraints shape accessible many-body information.


\begin{acknowledgments}
We thank Dr.~Chang Liu for helpful discussions. J.-X. Zhong was supported by the National Natural Science Foundation of China (Grant Nos.\ 12374046 and 11874316), the Shanghai Science and Technology Innovation Action Plan (Grant No.\ 24LZ1400800), and the National Basic Research Program of China (Grant No.\ 2015CB921103). H.-Z. Li is supported by a China Scholarship Council Scholarship (Grant No.\ 202506890103).
\end{acknowledgments}
\appendix

\section{Numerical evaluation of ensemble diagnostics}
\label{app:numerics}
\subsection{Moment evaluation from a wave function}
A normalized state can be arranged as a tensor $\Psi_{aez}$ in the chosen measurement basis. For each outcome, form the unnormalized target matrix
$\sigma_z=\Psi_z\Psi_z^\dagger, p_z=\tr\sigma_z,$
where $\Psi_z$ is a $d\times e$ matrix. Then
\begin{equation}
 \rho_A=\sum_z\sigma_z,
 \qquad
 \cM_{A,\mathrm P}^{(2)}=
 \sum_{z:p_z>0}\frac{\sigma_z\otimes\sigma_z}{p_z}.
 \label{eq:computemoment}
\end{equation}
For a full ensemble, set $e=1$. The general $k$th moment uses $\sigma_z^{\otimes k}/p_z^{k-1}$. The Born probabilities must appear to the first power after normalization; replacing them by uniform weights or by $p_z^2$ changes the quantity.

The matrices in Eq.~\eqref{eq:computemoment} are Hermitian. Their trace distances can be evaluated by summing the absolute eigenvalues of the Hermitian difference, after removing only roundoff-level anti-Hermitian components. The exact consistency identities are $\sum_zp_z=1$, $\tr\rho_A=1$, $\tr\cM^{(k)}=1$, $\tr C_A^{(2)}=0$, and agreement between the variance and swap-contraction expressions for $V$ in Sec.~\ref{sec:framework}. The variance bounds $V/2\leq\Delta\leq dV/2$ and Eqs.~\eqref{eq:infobounds} and \eqref{eq:contraction} provide further nontrivial checks. The first moment is also invariant under a complete change of measurement basis.

Exactly zero-probability outcomes are omitted. Nonzero outcomes contribute with their Born weights, including rare records. The roundoff scale near $10^{-16}$ visible in some trajectories limits the resolution of very small correlations.

\subsection{Long-time evolution and spectral precision}
For a finite Floquet system, a spectral representation allows evaluation at widely separated times without iterating every period. If $U_F=\sum_\mu\ee^{-\ii\theta_\mu}\ket{\mu}\bra{\mu}$, then
\begin{equation}
 \ket{\Psi_n}=\sum_\mu
       \ee^{-\ii n\theta_\mu}\ket{\mu}\langle\mu|\Psi_0\rangle.
 \label{eq:spectralevolution}
\end{equation}
This spectral method is used for the full projected ensembles; PPE and mutual-information trajectories are evolved period by period. At very long times, accuracy of the eigenphases becomes important: a phase error $\delta\theta_\mu$ accumulates as $n\delta\theta_\mu$. For illustration, an uncertainty of $10^{-15}$ multiplied by $10^{12}$ gives a phase uncertainty of order $10^{-3}$. This error affects phase-sensitive observables while preserving normalization, so spectral precision and state normalization characterize different aspects of numerical accuracy.

The eigenphase ordering, spacing normalization, entropy convention, and averaging windows used for the figures are specified in Appx.~\ref{app:provenance}.


\section{Numerical conventions}
\label{app:provenance}
The numerical calculations use the open-boundary Floquet model and random-product-state sampling specified in Sec.~\ref{sec:framework}. The full projected-ensemble reference is constructed separately from the Floquet diagonal ensemble of each initial state. Distances and mutual information are calculated before averaging over initial states. The exact ensemble identities in this paper hold for each state separately.

For Fig.~\ref{fig:regimes}(a), the half-chain von Neumann entropy is averaged over all $2^L$ Floquet eigenstates and divided by $L$, with natural logarithms and $L=8,10,12,14$. The control parameter is sampled at 24 equally spaced values in $[0,1]$. The displayed rescaling uses $F_c=0.651\pm0.005$ and $\nu=0.84\pm0.09$, obtained from variations across fitting windows and size sets. The windows are $[0.48,0.78]$, $[0.50,0.77]$, $[0.52,0.76]$, and $[0.54,0.74]$; the size sets are $\{8,10,12,14\}$, $\{10,12,14\}$, and $\{8,10,12\}$.

For the level-spacing distributions, $L=12$. Floquet eigenphases are sorted on $[0,2\pi)$, including the circular gap between the last and first phases. Each gap is divided by the mean gap, and the histogram uses 40 bins on $[0,4]$.

Full-ensemble residual distances use the last 200 saved times. In the ergodic PPE and mutual-information panels, the displayed residual signal is evaluated at $t=20$. In the constrained regime, both the PPE residual and the
mutual-information average are evaluated in the late-time
regime where the signals have settled, using the last 200
and 20 saved times, respectively.

The onset contours use an absolute threshold of $10^{-5}$. Their fitted coefficients can depend on interpolation between sampled times, the threshold, and the finite range of buffer lengths. The fits describe onset over the sampled buffer lengths and are compared with the effective-envelope model in Sec.~\ref{sec:mechanism}.


\section{Reference ensembles and moment constraints}
\label{app:full}
\subsection{Moment consistency and the first-moment floor}
For any ensemble of normalized states, tracing out one copy of its $k$th moment gives its $(k-1)$th moment. The same is true of a hierarchy generated from a fixed reference ensemble. Hence, for $j<k$,
\begin{equation}
 D_j=\frac12\norm{\tr_{k-j}\big(\cM^{(k)}-\cR^{(k)}\big)}_1
       \leq D_k.
 \label{eq:contraction}
\end{equation}
In particular, a Haar-design error is at least $\tfrac12\norm{\rho_A-\id/d}_1$. This is a pointwise relation and remains true after averaging the distances. The same averaging operation is applied to both distances in this inequality.

If the reference depends on $t$ through $\rho_A(t)$, its moments must still be generated consistently at that time. A reference matched to the instantaneous mean isolates higher-moment structure, whereas the Haar reference additionally tests proximity of the mean to $\id/d$.

\subsection{Higher-moment structure at a maximally mixed mean}
Consider pure qubit states
$\ket{\psi(\varphi)}=(\ket0+\ee^{\ii\varphi}\ket1)/\sqrt2$, with $\varphi$ uniformly distributed on $[0,2\pi)$. The mean density matrix is $\id/2$, but
$\cM^{(2)}=\frac{\id\otimes\id}{4} +\frac{X\otimes X+Y\otimes Y}{8}.$
The qubit Haar moment is $(\id+\Sswap)/6$. Their difference has eigenvalues $-1/12,-1/12,1/6,0$, so
\begin{equation}
 D_{1,\mathrm H}=0,\qquad D_{2,\mathrm H}=\frac16.
 \label{eq:equator}
\end{equation}
Uniform phase sampling at fixed amplitudes yields a maximally mixed average while retaining anisotropic second-moment structure. This separates first-moment relaxation from the formation of higher-order designs.

\subsection{Finite support and a matched-mean reference}
A full projected ensemble with $N$ outcomes of nonzero probability has a $k$th moment of rank at most $N$. If $D_{\rm sym}=\binom{d+k-1}{k}$, the support-projector measurement gives the lower bound
\begin{equation}
 D_{k,\mathrm H}\geq\max\left(0,1-\frac{N}{D_{\rm sym}}\right).
 \label{eq:rankbound}
\end{equation}
The bound is useful only when $N<D_{\rm sym}$, and becomes trivial once the number of outcomes is sufficiently large. It shows that finite ensemble size is a separate obstruction from retained dynamical memory.

One possible reference at a nonmaximally mixed mean is the Scrooge ensemble, whose construction minimizes accessible information at a prescribed mean. Let $\dd\mu_{\mathrm H}(x)$ be normalized Haar measure on unit vectors in dimension $d$, and use the reweighted probability density
$q(x)=d\bra{x}\rho\ket{x}.$
Map each sampled vector to
$\ket{\phi_x}=\frac{\sqrt\rho\ket{x}}{\sqrt{\bra{x}\rho\ket{x}}}.$
Then
\begin{equation}
 \int\dd\mu_{\mathrm H}(x)\,q(x)\ket{\phi_x}\bra{\phi_x}
 =d\sqrt\rho\left(\frac{\id}{d}\right)\sqrt\rho=\rho.
\end{equation}
The reweighting and normalized filter together enforce the prescribed first moment. This defines the reference independently of the dynamical relaxation toward it. In the full-ensemble calculations, the prescribed mean is the reduced Floquet diagonal ensemble defined in Sec.~\ref{sec:framework}.


\section{Covariance identities and information bounds}
\label{app:cov}
The statements in this appendix apply to any finite ensemble of density matrices $\{p_z,\rho_z\}$, including PPEs constructed from mixed global states. Let $\rho=\sum_zp_z\rho_z$, $\delta_z=\rho_z-\rho$, and $C=\sum_zp_z\rho_z\otimes\rho_z-\rho\otimes\rho$. Since $\sum_zp_z\delta_z=0$, expanding the tensor product gives
\begin{equation}
 C=\sum_zp_z\delta_z\otimes\delta_z.
 \label{eq:appcov}
\end{equation}
For Hermitian observables $O,O'$, define $m_z(O)=\tr(O\rho_z)$. Then
\begin{align}
 \tr[(O\otimes O')C]
 &=\sum_zp_z\big[m_z(O)-\overline m(O)\big]\nonumber\\
 &\quad\times\big[m_z(O')-\overline m(O')\big].
 \label{eq:classicalcov}
\end{align}
In particular, every same-observable covariance is nonnegative. This property does not imply that $C$ is positive as a two-copy operator: $\tr C=0$, and its nonzero eigenvalues necessarily include both signs.

The swap identity $\tr[\Sswap(X\otimes Y)]=\tr(XY)$ yields
\begin{equation}
 \tr(\Sswap C)=\sum_zp_z\norm{\delta_z}_2^2=V.
\end{equation}
Trace-norm duality and $\norm{\Sswap}_\infty=1$ imply $V\leq\norm C_1=2\Delta$. Conversely,
\begin{align}
 2\Delta
 &\leq\sum_zp_z\norm{\delta_z\otimes\delta_z}_1\nonumber\\
 &=\sum_zp_z\norm{\delta_z}_1^2
 \leq d\sum_zp_z\norm{\delta_z}_2^2=dV.
 \label{eq:proofbounds}
\end{align}
The last inequality is the Cauchy--Schwarz bound on the singular values of a $d\times d$ matrix. Since every summand in $V$ is nonnegative, $\Delta=0$ holds exactly when $\rho_z=\rho$ for all $p_z>0$.

For a classical register $Z$ carrying the label $z$, define $\omega_{AZ}=\sum_zp_z\rho_z\otimes\ket{z}\bra{z}$. Its mutual information is
$I(A{:}Z)_\omega=\sum_zp_zD(\rho_z\Vert\rho)=\chi,$
where $D(\sigma\Vert\rho)=\tr[\sigma(\ln\sigma-\ln\rho)]$. The support condition is automatically satisfied for every outcome of nonzero weight. Quantum Pinsker's inequality~\cite{Audenaert2014}, in natural-log units, gives
$D(\rho_z\Vert\rho)\geq\frac12\norm{\rho_z-\rho}_1^2.$
Combining it with the first two steps of Eq.~\eqref{eq:proofbounds} proves $\Delta\leq\chi$. By the data-processing inequality~\cite{Lindblad1975}, a measurement channel $B\rightarrow Z$ cannot increase quantum mutual information, giving the information bounds
\begin{equation}
 \Delta\leq\chi\leq I_q(A{:}B).
 \label{eq:appendixinfobounds}
\end{equation}
When entropy is expressed in bits, the first inequality becomes $\Delta\leq(\ln2)\chi_{\rm bits}$.

These bounds are not equalities in general. For example,
$\rho_{AB}=\frac14(\id\otimes\id+cX\otimes X)$, with $0<|c|\leq1$,
has nonzero quantum mutual information. Measuring $B$ in the $Z$ basis nevertheless yields $\rho_{A|z}=\id/2$ for both outcomes, so $\Delta=\chi=0$. The measurement basis selects which correlations are retained by the PPE.

Finally, consider a product-state quench in a strictly finite-depth local circuit. If the backward causal regions of $A$ and $B$ are disjoint, all joint expectation values factorize and $\rho_{AB}=\rho_A\otimes\rho_B$. Hence $I_q=\chi=\Delta=0$. This is an exact kinematic statement under the stated circuit and initial-state assumptions. It is distinct from an effective slow front within the microscopic causal region, where correlations may be nonzero but extremely small. None of these statements implies signalling by the final measurement: its unconditioned action leaves $\rho_A$ unchanged.


\section{Supporting results}
\label{app:supporting}
\begin{figure}[!tbp]

\centering
\includegraphics[width=\columnwidth]{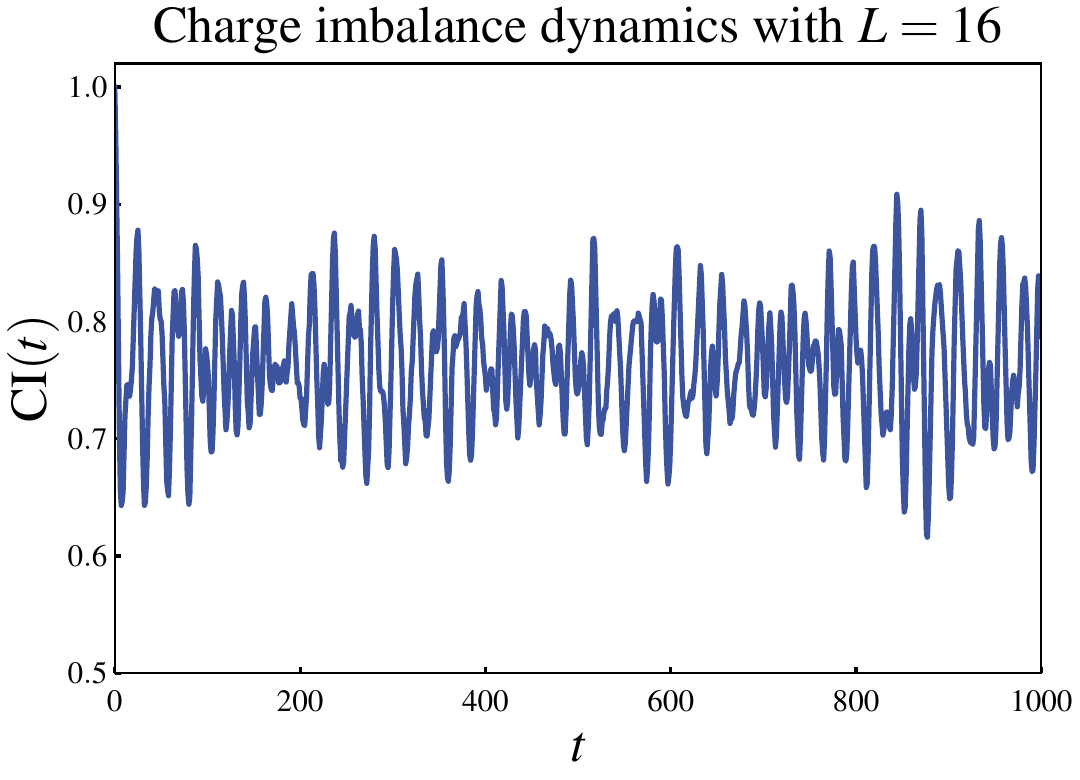}
\caption{Local memory at $F=0.8$ and $L=16$. The staggered magnetization $\mathrm{CI}(t)=L^{-1}\sum_{j=0}^{L-1}(-1)^j\langle Z_j(t)\rangle$ starts from the N\'eel state $\ket{0101\cdots}$, with $\mathrm{CI}(0)=1$. Time is in Floquet periods.}
\label{fig:memory}

\end{figure}

Fig.~\ref{fig:memory} shows the staggered magnetization following a N\'eel-state quench at $F=0.8$ and $L=16$. Its large positive value and persistent oscillations indicate local memory over the displayed interval.

\begin{figure}[!tbp]

\centering
\includegraphics[width=\columnwidth]{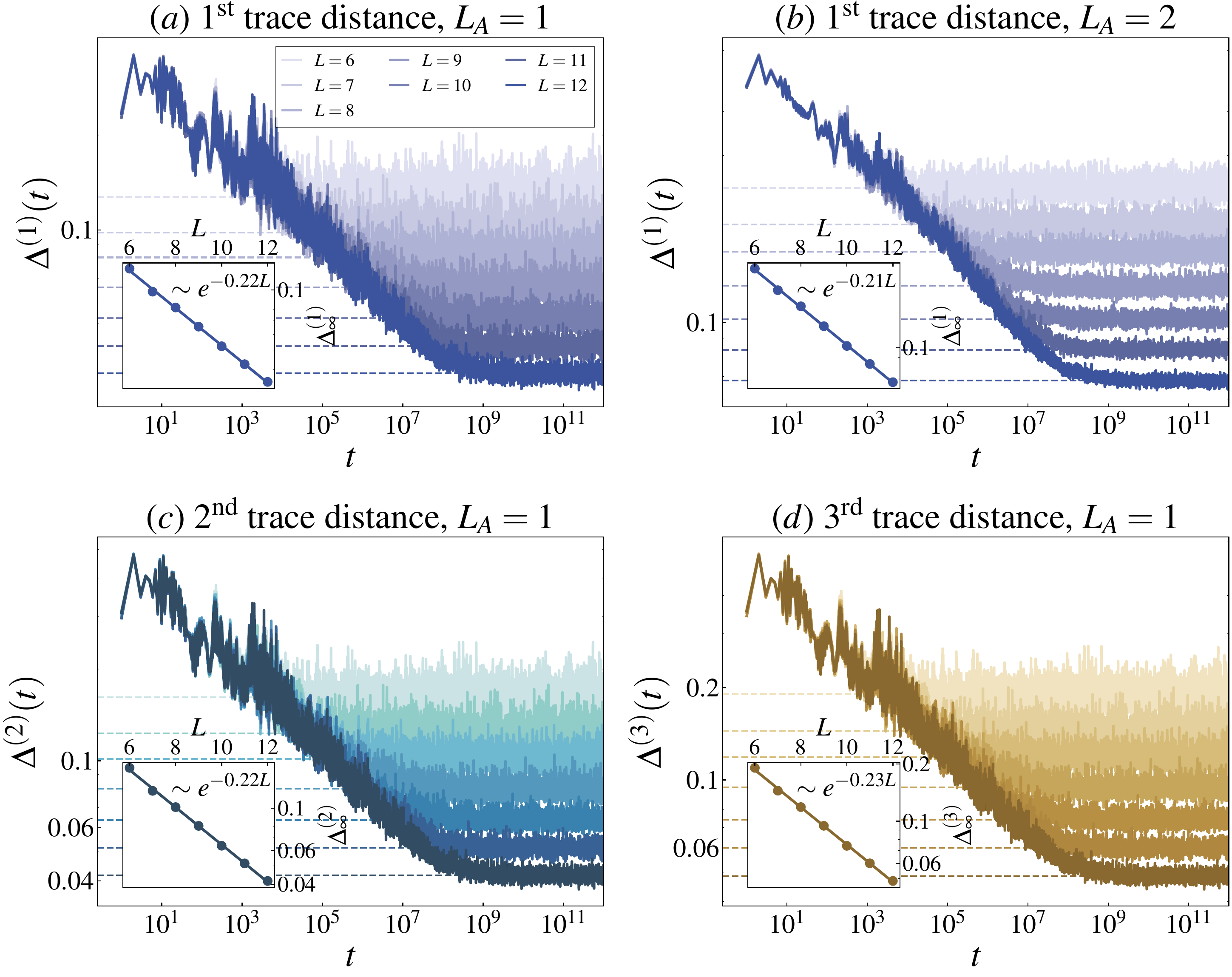}
\caption{Full-ensemble moment dependence at $F=0.8$ with $X$-basis measurements, averaged over 1000 random product states. The trace distances $\Delta^{(k)}$ use the initial-state-specific Scrooge reference. Panels (a)--(d) have $(L_A,k)=(1,1),(2,1),(1,2),(1,3)$, respectively, for $L=6$--$12$. Horizontal dashed lines and inset points are averages over the last 200 saved times; inset lines are exponential fits with the indicated exponents. Time is in Floquet periods.}
\label{fig:additional}

\end{figure}

Fig.~\ref{fig:additional} supplements the full-ensemble analysis with the same Scrooge reference as in Sec.~\ref{sec:framework}. The first moments for $L_A=1,2$ show slow relaxation and decreasing residual distances with system size, with exponential slopes $0.22$ and $0.21$. The second and third moments for $L_A=1$ give slopes $0.22$ and $0.23$. These comparable size dependences are consistent with the shared first-moment contribution to the higher-moment distances established in Appx.~\ref{app:full}.

\section{Induced-state benchmarks and a finite recorded region}
\label{app:random}
Random induced states provide a useful reference for the loss of accessible fluctuations under a partial trace. The Page entropy~\cite{Page1993} is a related benchmark for the entanglement of random pure states; neither reference fixes a dynamical onset time. Let $\ket{\psi}$ be Haar distributed on a tensor product of dimensions $d$ and $e$, and define $\rho_\psi=\tr_E\ket{\psi}\bra{\psi}$. The Haar two-copy identity gives
$\mathbb{E}\big[(\ket{\psi}\bra{\psi})^{\otimes2}\big] =\frac{\id+\Sswap_A\otimes\Sswap_E}{de(de+1)}.$
Tracing both environment copies yields
\begin{equation}
 \mathbb{E}\rho_\psi^{\otimes2}
   =\frac{e\id+\Sswap_A}{d(de+1)},
 \qquad \mathbb{E}\rho_\psi=\frac{\id}{d}.
 \label{eq:indmoment}
\end{equation}
Thus the connected covariance of the continuous induced ensemble is
\begin{equation}
 C_{\mathrm{ind}}=
 \frac{\Sswap_A-\id/d}{d(de+1)}.
 \label{eq:indC}
\end{equation}
The swap has eigenvalues $+1$ and $-1$ in subspaces of dimensions $d(d+1)/2$ and $d(d-1)/2$. Summing the absolute eigenvalues of Eq.~\eqref{eq:indC} gives Eq.~\eqref{eq:inducedmain}. Its variance is $V_{\mathrm{ind}}=(d^2-1)/[d(de+1)]$, so the lower bound $\Delta\geq V/2$ is saturated.

A finite recorded region changes the ensemble-mean connected covariance. Let a single global state on $AEB$ be Haar distributed, with dimensions $d,e,b$, and measure $B$ in a fixed orthonormal basis. The normalized conditional vectors on $AE$ are independent Haar vectors whose directions are independent of their Born weights. Consequently, the average PPE second moment is still Eq.~\eqref{eq:indmoment}. However, the mean state of each \emph{realization} is $\rho_A$, not identically $\id/d$. Its second tensor moment is
$\mathbb{E}\rho_A^{\otimes2} =\frac{eb\id+\Sswap_A}{d(deb+1)}.$
Subtracting gives the exact ensemble-mean connected covariance
\begin{align}
 \mathbb{E}C_A^{(2)}&=\gamma\big(\Sswap_A-\id/d\big),\\
 \gamma&=\frac{e(b-1)}{(de+1)(deb+1)}.
 \label{eq:finiteB}
\end{align}
In particular,
\begin{equation}
 \frac12\norm{\mathbb{E}C_A^{(2)}}_1
 =\frac{e(b-1)(d^2-1)}{2(de+1)(deb+1)}
 =\frac{\mathbb{E}V}{2}.
 \label{eq:finiteBnorm}
\end{equation}
The right-hand side is a lower bound on $\mathbb{E}\Delta$, not generally an equality. Interchanging an average with a trace norm would discard sample fluctuations. At $b=1$ the PPE consists of one state and its connected covariance vanishes exactly. As $b\rightarrow\infty$, Eq.~\eqref{eq:finiteBnorm} approaches the continuous induced-ensemble result. These limits show why varying $L_B$ along with $L_E$ can affect an apparent attenuation law.

These reference formulas describe attenuation under a partial trace. They contain no time scale and therefore separate the effect of unobserved Hilbert-space dimension on signal amplitude from the dynamics of correlation onset.


\section{Exact solution of a diagonal dephasing model}
\label{app:dephasing}
\subsection{Construction of the effective diagonal description}
The starting point is the Floquet operator $U_F=\ee^{-\ii K}U_0$, with $U_0=\ee^{-\ii H_z}$ and $K=g\sum_jX_j$. At $g=0$, computational-basis configurations $\ket{s}$ are eigenstates of $U_0$, with eigenvalues $d_s=\ee^{-\ii E_s}$. A nonresonant kick can be removed perturbatively by a unitary change of basis. Writing $U_F=U_0+\delta U+O(g^2)$, where $\delta U=-\ii K U_0$, choose an anti-Hermitian generator $S$ so that $\ee^S U_F\ee^{-S}$ is diagonal to first order. Its off-diagonal entries vanish when
\begin{equation}
 S_{ss'}=\frac{\ii K_{ss'}d_{s'}}{d_{s'}-d_s},\qquad s\ne s'.
 \label{eq:dressinggenerator}
\end{equation}
This expansion requires the relevant kick matrix elements to be small compared with the quasienergy denominators. Resonant configurations must instead be treated together as blocks. Iterating the nonresonant elimination motivates a dressed-spin description when the resulting transformation remains quasi-local on the scales of interest.

Let $\mathcal U$ denote such a dressing transformation, with $\mathcal U^\dagger U_F\mathcal U=\exp[-\ii f(\{Z_j\})]$, and define $\tau_j^z=\mathcal U Z_j\mathcal U^\dagger$. On a chosen quasienergy branch, the diagonal generator has the expansion
\begin{equation}
 \begin{aligned}
 H_{\mathrm{eff}}&=\epsilon_0+\sum_i h_i\tau_i^z
 +\sum_{i<j}J_{ij}\tau_i^z\tau_j^z\\
 &\quad+\sum_{i<j<k}J_{ijk}\tau_i^z\tau_j^z\tau_k^z+\cdots.
 \end{aligned}
 \label{eq:effectiveexpansion}
\end{equation}
For a fixed eigenstate labeling and branch, each coefficient is the discrete expansion coefficient $J_{\mathcal S}=2^{-L}\sum_{\boldsymbol s}f(\boldsymbol s)\prod_{j\in\mathcal S}s_j$. This identity specifies how diagonal many-spin terms arise; their spatial decay additionally requires quasi-local dressing. The scalar $\epsilon_0$ has no effect on conditional states. Retaining the one-spin and pairwise terms gives the dephasing Hamiltonian
\begin{equation}
 H_{\mathrm d}=\sum_i h_i\tau_i^z
          +\sum_{i<j}J_{ij}\tau_i^z\tau_j^z,
 \label{eq:diagonal}
\end{equation}
The $h_i$ and $J_{ij}$ here are dressed coefficients, not the bare field profile and nearest-neighbor Ising coupling. The pairwise truncation is a minimal effective model, and the long-distance envelope of its couplings is an additional input. For an approximate construction, a remainder $R$ in the effective generator limits its useful time window; a sufficient finite-system condition is $t\|R\|\ll1$. The effective description is used subject to these nonresonance and truncation conditions.

Transforming the physical problem also dresses its initial state and measurement operators. The exact calculation below adopts a product state and measurements diagonal in the effective-spin basis to isolate conditional dephasing. A physical product state and an $X$-basis measurement need not have those forms after dressing. Thus the construction explains the origin and truncation of the model used in Sec.~\ref{sec:mechanism}, while its factorized solution isolates a possible dephasing mechanism. The relation to general and physical measurements is specified in Appx.~\ref{app:generalmeasurement}.

\subsection{Conditional states and connected moments}
Let $A$ contain one effective spin, with eigenvalues $s=\pm1$ of $\tau_A^z$. Every initial spin has amplitudes $a_j,b_j$ in the $\tau_j^z$ basis, with $|a_j|^2=q_j$ and $|b_j|^2=1-q_j$. The initial target coherence is $c_0=a_A b_A^*$. For fixed configurations $\eta$ on $E$ and $z$ on $B$, the diagonal Hamiltonian in Eq.~\eqref{eq:diagonal} can be written
\begin{align}
 E(s,\eta,z)&=s\left[h_A+\sum_{j\in E}J_{Aj}\eta_j
                          +\sum_{j\in B}J_{Aj}z_j\right]\nonumber\\
 &\quad+E_{EB}(\eta,z).
\end{align}
The probability of a $B$ outcome is the time-independent product probability $p_z=\prod_{j\in B}q_j^{(1+z_j)/2}(1-q_j)^{(1-z_j)/2}$. In the off-diagonal target matrix element, the phase generated by $E_{EB}$ cancels between bra and ket. Summing over $\eta$ gives
\begin{equation}
 c_z(t)=c_0\ee^{-2\ii h_At}G_E(t)
               \ee^{-2\ii t\sum_{j\in B}J_{Aj}z_j}.
 \label{eq:condcoherence}
\end{equation}
Here $G_X^{(n)}=\prod_{j\in X}[q_j\ee^{-2\ii nJ_{Aj}t}+(1-q_j)\ee^{2\ii nJ_{Aj}t}]$ and $G_X=G_X^{(1)}$. Thus
\begin{align}
 \rho_{A|z}(t)&=\begin{pmatrix}q_A&c_z(t)\\c_z(t)^*&1-q_A\end{pmatrix},\nonumber\\
 \rho_A(t)&=\begin{pmatrix}q_A&c(t)G_B(t)\\c(t)^*G_B(t)^*&1-q_A\end{pmatrix}.
 \label{eq:dephasingstates}
\end{align}
where $c(t)=c_0\ee^{-2\ii h_At}G_E(t)$. Arbitrary pairwise diagonal couplings within $E\cup B$ are included in $E_{EB}$ and do not affect this result. Multispin terms involving $A$ generally produce a nonfactorized conditional field and need a separate treatment.

Write $u_z=c_z-cG_B$. The fluctuation matrix has only off-diagonal entries, $\delta_z=\bigl(\begin{smallmatrix}0&u_z\\u_z^*&0\end{smallmatrix}\bigr)$. Define $\ell=\sum_zp_z|u_z|^2=|c|^2(1-|G_B|^2)$ and $k=\sum_zp_zu_z^2=c^2\left(G_B^{(2)}-G_B^2\right)$.
In the ordered two-copy basis $\{++,+-,-+,--\}$,
\begin{equation}
 C_A^{(2)}=\begin{pmatrix}
 0&0&0&k\\
 0&0&\ell&0\\
 0&\ell&0&0\\
 k^*&0&0&0
 \end{pmatrix}.
 \label{eq:matrixC}
\end{equation}
Its eigenvalues are $\{\ell,-\ell,|k|,-|k|\}$. Therefore $V=2\ell$ and $\Delta=\ell+|k|$, giving
\begin{equation}
 V(t)=2|c_0|^2|G_E(t)|^2
                 \left[1-|G_B(t)|^2\right].
 \label{eq:Vdephase}
\end{equation}
\begin{align}
 \Delta(t)&=|c_0|^2|G_E(t)|^2\,\mathcal{F}_B(t),
 \label{eq:exactDelta}\\
 \mathcal{F}_B(t)&=1-|G_B(t)|^2
       +\left|G_B^{(2)}(t)-G_B(t)^2\right|.
 \label{eq:FB}
\end{align} The bound $|k|\leq\ell$, following from $|\sum p_zu_z^2|\leq\sum p_z|u_z|^2$, also reproduces $V/2\leq\Delta\leq V$ for a qubit.

When $|c|>0$, the buffer-independent normalized variance is
\begin{equation}
 \mathcal{R}_B(t)=
 \frac{V(t)}{2\sum_zp_z|[\rho_{A|z}(t)]_{+-}|^2}
 =1-|G_B(t)|^2.
 \label{eq:normalizedR}
\end{equation}
The normalization isolates remote phase fluctuations whenever the conditional coherence is nonzero and resolved. Its interpretation uses the common coherence magnitude of this model and measurement basis.

\subsection{Holevo information and early-time response}
All conditional states share the same eigenvalues. With $m=2q_A-1$, define
$r_c=\sqrt{m^2+4|c|^2}, r_m=\sqrt{m^2+4|cG_B|^2}.$
Writing $h_2(x)=-x\ln x-(1-x)\ln(1-x)$, the exact record information is
\begin{equation}
 \chi=h_2\!\left(\frac{1+r_m}{2}\right)
       -h_2\!\left(\frac{1+r_c}{2}\right).
 \label{eq:exactHolevo}
\end{equation}
For a balanced target and small surviving coherence, the binary-entropy expansion gives
\begin{equation}
 \chi=2|c|^2(1-|G_B|^2)+O(|c|^4)=V+O(|c|^4).
\end{equation}
Thus PPE fluctuations and record information respond to the same conditional phases in this regime, although their magnitudes and finite-threshold onset times need not coincide.

Since $G_B=\langle\ee^{-2\ii tH_B}\rangle$, the cumulant expansion gives $|G_B|^2=1-4t^2\operatorname{Var}(H_B)+O(t^4)$. Independent initial spins yield
\begin{equation}
 \sigma_B^2=4\sum_{j\in B}q_j(1-q_j)J_{Aj}^{\,2}.
 \label{eq:sigmab}
\end{equation} The characteristic onset scale is therefore set by the variance of a \emph{conditional} field, not by a bare local precession frequency $h_A$; the latter cancels from $\Delta$, $V$, and $\chi$.

\subsection{Time-averaged buffer coherence}
For a single buffer spin,
$|G_j|^2=q_j^2+(1-q_j)^2 +2q_j(1-q_j)\cos(4J_{Aj}t).$
If no nontrivial frequency combination appearing in the finite product has zero frequency, its infinite-time average is the product of the constant terms. This gives
\begin{equation}
 \overline{|G_E|^2}
    =\prod_{j\in E}\left[q_j^2+(1-q_j)^2\right].
 \label{eq:bufferplateau}
\end{equation} The condition can fail for commensurate or degenerate couplings. Even when it holds, a finite system has recurrences and the statement concerns a time average, not pointwise convergence. Replacing $\mathcal{F}_B$ by an order-one constant additionally requires enough independently dephased measured spins; for one or a few recorded spins, its oscillations and second harmonic must be retained explicitly. The product determines the attenuation scale, while finite-record oscillations and correlations between averaging factors determine its prefactor.

\subsection{General measurement bases and physical projectors}
\label{app:generalmeasurement}
The role of the measurement basis can be made explicit without assuming a product initial state. In the tensor-product coordinates of a diagonal model, let $\Psi_0(s,\eta,b)$ be the initial wavefunction and $E(s,\eta,b)$ its configuration energy. For a measurement basis on $B$, write $m_z(b)=\langle z;\alpha|b\rangle$. The unnormalized conditional amplitudes and target state are
\begin{align}
 \Phi_z(s,\eta;t)&=\sum_b m_z(b)\ee^{-\ii t E(s,\eta,b)}
                  \Psi_0(s,\eta,b),\label{eq:generalbasisamplitude}\\
 [\sigma_{A,z}(t)]_{ss'}&=\sum_\eta\Phi_z(s,\eta;t)
                                  \Phi_z(s',\eta;t)^*.
 \label{eq:generalbasisstate}
\end{align}
Here $p_z=\operatorname{Tr}\sigma_{A,z}$ and $\rho_{A|z}=\sigma_{A,z}/p_z$ for $p_z>0$. Longitudinal measurement gives $m_z(b)=\delta_{zb}$ and recovers the preceding solution for a product initial state. For transverse measurement in these coordinates, $m_z(b)=2^{-L_B/2}\prod_{j\in B}z_j^{(1-b_j)/2}$, with $z_j,b_j=\pm1$. Different configurations then interfere within each outcome. Conditional populations, coherence magnitudes, and outcome probabilities can all change; phases from interactions involving measured spins within $E\cup B$ need not cancel. Consequently, the product formulas for $V$ and $\Delta$ are specific to the longitudinal-measurement setting.

Physical $X$ measurements require one further step because dressing need not preserve the physical partition. Set $D=f(\{Z_j\})$, $\widetilde\rho_0=\mathcal U^\dagger\rho_0\mathcal U$, and $\widetilde P_z=\mathcal U^\dagger P_z^X\mathcal U$, where $P_z^X$ acts as the identity on $AE$. Writing $\widetilde\rho(t)=\ee^{-\ii tD}\widetilde\rho_0\ee^{\ii tD}$, the physical conditional state is exactly
\begin{equation}
 \sigma_{A,z}^X(t)=\operatorname{Tr}_{EB}\!\left[
 \mathcal U\widetilde P_z\widetilde\rho(t)
 \widetilde P_z\mathcal U^\dagger\right].
 \label{eq:dressedphysicalmeasurement}
\end{equation}
This identity uses the full diagonal representation; a truncated generator introduces the approximation described above. It retains both the dressed preparation and projectors and takes the partial trace in the physical partition. Eqs.~\eqref{eq:generalbasisamplitude} and \eqref{eq:generalbasisstate} apply directly when the projectors act only on $B$ in the chosen tensor-product coordinates. Eq.~\eqref{eq:dressedphysicalmeasurement} does not require that restriction. The covariance and information bounds remain valid for the resulting physical ensemble, whereas transferring the factorized response or its onset law requires additional microscopic justification.


\section{Spatial onset laws and threshold dependence}
\label{app:fronts}
\subsection{Effective coupling and onset scale}
The conditional-field argument in Sec.~\ref{sec:mechanism} identifies the normalized onset scale with
\begin{equation}
 t_*(r)J_{\mathrm{eff}}(r)\sim1,
 \qquad r=L_E+O(1),
 \label{eq:onsetcoupling}
\end{equation}
A static virtual path with successive detunings $n\mathcal E$ provides the illustrative amplitude
\begin{equation}
 A_r\sim g\prod_{n=1}^{r-1}\frac{g}{n\mathcal{E}}
       =g\frac{(g/\mathcal{E})^{r-1}}{(r-1)!}.
 \label{eq:virtualpath}
\end{equation}
The corresponding effective-envelope hypothesis is
\begin{equation}
 J_{\mathrm{eff}}(r)=J_0\exp\!\left[-ar\ln r-b_1 r+o(r)\right],
 \qquad a>0.
 \label{eq:tail}
\end{equation}
For a Floquet system, the perturbative denominators instead obey
\begin{equation}
 \left|\ee^{-\ii T\varepsilon_m}-\ee^{-\ii T\varepsilon_n}\right|
 =2\left|\sin\frac{T(\varepsilon_m-\varepsilon_n)}{2}\right|.
 \label{eq:Floquetdenominator}
\end{equation}
The accumulating-detuning argument applies within an appropriate nonresonant small-phase window. Under the envelope hypothesis, the resulting onset law is
\begin{equation}
 \ln\frac{t_*(r)}{t_0}=ar\ln r+b_1 r+o(r).
 \label{eq:frontlaw}
\end{equation}

Ignoring the $o(r)$ term in Eq.~\eqref{eq:frontlaw}, set $y=\ln(t/t_0)$ and $u=\ln r+b_1/a$. Then $r=\ee^{u-b_1/a}$ and
$u\ee^u=\frac{y}{a}\ee^{b_1/a}.$
Selecting the principal Lambert branch gives
\begin{equation}
 r(t)=\frac{y}{a\,W_0\!\left[(y/a)\ee^{b_1/a}\right]},
 \label{eq:lambert}
\end{equation} The asymptotic expansion $W_0(x)=\ln x-\ln\ln x+o(1)$ gives
\begin{equation}
 r(t)\sim\frac{\ln(t/t_0)}{a\ln\ln(t/t_0)}.
 \label{eq:sublog}
\end{equation} The constants $t_0$ and $b_1$ enter slowly varying corrections, which can be substantial over the available range. Throughout the front analysis, $b_1$ denotes the linear-in-distance coefficient; it is distinct from the field curvature $\beta$ and the measured-region dimension $b$.

A fixed offset $r=L_E+r_0$ also matters at small separation. Expanding $(L_E+r_0)\ln(L_E+r_0)$ produces terms of order $\ln L_E$ in addition to the leading $L_E\ln L_E$. Such corrections can shift fitted prefactors without changing a genuine large-distance leading coefficient. They contribute to finite-distance fits over short buffers.

\subsection{Threshold dependence and spatial diagnostics}
Suppose an observable has the approximate form
\begin{equation}
 Q(t,r)=A(r)f\!\left(t/\tau(r)\right),
 \qquad f(\infty)=1.
 \label{eq:scalingansatz}
\end{equation}
A fixed absolute threshold $Q=\epsilon$ gives
\begin{equation}
 t_\epsilon(r)=\tau(r)f^{-1}\!\left(\frac{\epsilon}{A(r)}\right).
 \label{eq:absthreshold}
\end{equation}
If $A(r)$ decreases, this time can increase even at fixed $\tau(r)$. It ceases to exist when the threshold is above the accessible signal. For an early rise $f(x)\simeq Cx^p$ and $A(r)\sim\ee^{-sr}$, the threshold contributes an additional term $sr/p$ to $\ln t_\epsilon$. It can therefore renormalize a linear-in-distance term in a fit or exaggerate a finite-distance delay. Here $p$ characterizes the assumed early-time response.

A relative threshold $Q=qA(r)$, $0<q<1$, instead gives $t_q=\tau(r)f^{-1}(q)$. This normalization uses a resolved reference amplitude after baseline subtraction. For an oscillatory signal, the extracted onset additionally depends on the smoothing window or sustained-crossing rule.

The diagonal model provides an additional normalization through Eq.~\eqref{eq:normalizedR}, which does not require a late-time plateau. It applies when the common conditional coherence is nonzero and resolved. Comparing these normalizations separates phase accumulation from changes in signal amplitude.

Given onset data $t_*(r)$, the direct variable for model comparison is $y(r)=\ln t_*(r)$. Useful competing hypotheses include
\begin{align}
 y_{\rm exp}(r)&=c+\lambda r,\\
 y_{\rm fac}(r)&=c+ar\ln r,\\
 y_{\rm pow}(r)&=c+z\ln r.
 \label{eq:fitmodels}
\end{align}
Over short intervals, the $r\ln r$ and $r$ terms are strongly correlated. Distinguishing the factorial envelope from exponential or crossover behavior therefore depends on spatial range and uncertainty in the extracted onset times. Figs.~\ref{fig:ppe} and \ref{fig:mi} use the reduced factorial form as a finite-distance description.

A local curvature diagnostic is
$\kappa(r)=y(r+1)-2y(r)+y(r-1).$
For the factorial form, $\kappa(r)=a/r+O(r^{-3})$, whereas an exact exponential gives zero. The curvature cancels constant and linear terms, providing a diagnostic of the spatial envelope. As a second difference, it also amplifies uncertainty in individual onset times.

Finite-distance onset coefficients depend on the threshold, buffer interval, and recorded-region size. Uncertainty in a reference amplitude also propagates into normalized onsets. Statistical uncertainty across initial states is distinct from temporal fluctuations within a trajectory, whose saved times are correlated.


\section{Experimental implementation of partial projected ensembles}
\label{app:implementation}
The PPE protocol is compatible with digital quantum simulation~\cite{Feynman1982,Lloyd1996}, with separate costs for implementing the dynamics and estimating conditional observables. Projected-ensemble measurements have already been used to characterize many-body dynamics on a superconducting processor. For Eq.~\eqref{eq:floquet}, one period applies $R_z(2h_j)$ rotations and nearest-neighbor $\exp(-\ii JZ_jZ_{j+1})$ gates, followed by single-qubit $R_x(2g)$ rotations. Since the diagonal terms commute, this decomposition is exact for that Floquet unitary. Even and odd two-qubit bonds can be scheduled separately.

After preparing a chosen product state and evolving for a specified number of periods, the experiment measures $B$ in the desired basis and performs informationally complete measurements on the small subsystem $A$. Randomized measurement protocols~\cite{Elben2019,Brydges2019,Elben2023Toolbox} and classical shadows~\cite{Huang2020Shadows} provide candidate tools for estimating low-order observables, although a PPE estimator must also account for conditioning on $B$. Region $E$ is left unobserved, or its measured outcomes are discarded rather than conditioned upon. These two procedures produce the same PPE in the ideal circuit. Repeating the experiment gives the Born weights and the conditional observables needed to reconstruct $\rho_{A|z}$ or a selected covariance witness. Holding $L_A$ and $L_B$ fixed isolates changes in the buffer geometry; changing $L_B$ also changes the measurement ensemble.

The main practical obstacle is the size of the outcome register, not solely the size of $A$. Two independent experimental repetitions with the same outcome $z$ occur with probability $p_z^2$, whereas the PPE moment weights that outcome by $p_z$. A naive same-outcome pair average therefore estimates the wrong ensemble. Conditional estimation, appropriately corrected weights, or a separately validated estimator is required. Rare outcomes, nonlinear trace norms, and readout errors can all bias a reconstructed onset. The variance $V$ or covariances of a few target observables may be more accessible than full moment tomography, with sampling cost set by the outcome probabilities and target precision.

Three comparisons are especially diagnostic of the proposed mechanism. Comparing normalized and absolute-threshold onsets separates delay from loss of amplitude. Changing the populations of the buffer while keeping the remote conditional couplings comparable primarily affects the surviving signal in the diagonal picture. Changing the remote measurement basis tests whether an apparent absence of correlations is due to dynamics or to a record that is insensitive to them. These experimental comparisons target circuit depths accessible to the hardware; the long-time full-ensemble calculations instead use spectral propagation.

Error mitigation~\cite{Temme2017,LiBenjamin2017,Cai2023,Li2026Mitigation} may improve conditional-state reconstruction, with its effect on estimator bias and variance requiring validation. Comparisons with monitored or nonunitary dynamics, including processor realizations of the non-Hermitian skin effect~\cite{Shen2024Skin}, could distinguish information selected by a final record from changes induced by measurement backaction.


\emergencystretch=1em
\bibliographystyle{apsrev4-2-titles}
\bibliography{refs}

\begin{thebibliography}{90}
\bibitem [{Bas()}]{BasisConsistencyNote}%
  \BibitemOpen
  \href@noop {} {}\bibinfo {note} {For a complete projective measurement and Born-weighted averaging, the first projected-ensemble moment equals the reduced state in every measurement basis. Apparent basis dependence of this first moment indicates inconsistent weights, incomplete outcome coverage, or numerical error. Higher moments can remain basis dependent.}\BibitemShut {Stop}%
\bibitem [{Thr()}]{ThresholdOnsetNote}%
  \BibitemOpen
  \href@noop {} {}\bibinfo {note} {For a response $Q(t,r)=A(r)f[t/\tau(r)]$, a fixed absolute threshold $Q=\epsilon$ gives $t_\epsilon(r)=\tau(r)f^{-1}[\epsilon/A(r)]$. Attenuation can therefore delay the crossing even at fixed intrinsic time scale. A relative threshold $Q=qA(r)$ instead gives $t_q(r)=\tau(r)f^{-1}(q)$; different diagnostics need not have identical threshold-crossing times or fitted coefficients.}\BibitemShut {Stop}%
\end{thebibliography}
\end{document}